\documentclass[%
 aip,
 amsmath,amssymb,
 reprint,%
]{revtex4-1}

\usepackage{graphicx}
\usepackage{dcolumn}
\usepackage{bm}

\usepackage[utf8]{inputenc}
\usepackage[T1]{fontenc}
\usepackage{mathptmx}
\usepackage{etoolbox}
\usepackage{subcaption}

\makeatletter
\def\@email#1#2{%
 \endgroup
 \patchcmd{\titleblock@produce}
  {\frontmatter@RRAPformat}
  {\frontmatter@RRAPformat{\produce@RRAP{*#1\href{mailto:#2}{#2}}}\frontmatter@RRAPformat}
  {}{}
}%
\makeatother
\begin{document}

\preprint{AIP/123-QED}

\title{Quantum Kinetic Modeling of KEEN Waves in a Warm-Dense Regime}
\author{F. Alejandro Padilla-Gomez}
\email{padill77@msu.edu.}
\affiliation{Michigan State University, Department of Computational Mathematics, Science, and Engineering}
\author{Sining Gong}%
\affiliation{Michigan State University, Department of Computational Mathematics, Science, and Engineering}

\author{Michael S. Murillo}
\affiliation{Michigan State University, Department of Computational Mathematics, Science, and Engineering}

\author{F. R Graziani}
\affiliation{%
Lawrence Livermore National Laboratory, High Energy Density Science
}%

\author{Andrew J. Christlieb}
\affiliation{Michigan State University, Department of Computational Mathematics, Science, and Engineering}

\date{\today}

\begin{abstract}

We report the first fully kinetic, quantum study of Kinetic Electrostatic Electron Nonlinear (KEEN) waves, showing that quantum diffraction systematically erodes the classical trapping mechanism, narrow harmonic locking to the fundamental, and hasten post-drive decay. Electrons are evolved with a second-order Strang-split 1D1V Wigner–Poisson solver that couples conservative semi-Lagrangian WENO advection to an analytic Fourier space update for the non-local Wigner term, while ions remain classical. Short, frequency-tuned ponderomotive pulses drive KEEN formation in a uniform Maxwellian plasma; as the dimensionless quantum parameter H rises from the classical limit to values relevant to warm-dense matter, doped semiconductors, and 2D electron systems, the drive threshold increases, higher harmonics are damped, trapped electron vortices diffuse, and the subplasma electrostatic energy relaxes to a lower stationary level, as confirmed by continuous wavelet analysis.
These microscopic changes carry macroscopic weight. Ignition-scale capsules now compress matter to regimes where the electron de Broglie wavelength rivals the Debye length, making classical kinetic descriptions insufficient. By extending KEEN physics into this quantum domain, our results offer a potential diagnostic of nonequilibrium electron dynamics for next-generation inertial-confinement designs and high-energy-density platforms, indicating that predictive fusion modeling may benefit from the integration of kinetic fidelity with quantum effects.

\end{abstract}

\maketitle

\section{\label{sec:level1}Introduction}

We report the first fully kinetic, quantum extension of Kinetic Electrostatic Electron Nonlinear (KEEN) waves, demonstrating that quantum diffraction fundamentally reshape their excitation, nonlinear trapping, harmonic locking, and longevity. Using the Wigner-Poisson system, an approach inspired by Kinetic Theory Molecular Dynamics (KTMD) \cite{PhysRevE.90.033104}. We drive plasmas with short  pulses that seed subplasma frequency, multiharmonic electron responses with non-linear normal mode analog.  In the classical setting, this sets up the KEEN wave.  In this quantum regime, diffraction reorganizes trapped electrons in phase space vortices and modifies collisionless resonant phase mixing (Landau damping) by narrowing the resonance and altering trapping, kinetic features not present in fluid models that ultimately control KEEN persistence.

A broad class of quantum plasmas (i.e., warm-dense matter, doped semiconductors, 2D electron systems, etc.) host longitudinal plasmons that can be directly driven by optical, IR, or THz fields \cite{glenzer2007observations, fang2013gated}. In these settings, nonequilibrium driven electron distributions and their nonlinear dielectric response determine energy flow, wave-particle coupling, and stopping power for energetic projectiles\cite{malko2022proton, cayzac2017experimental}. Motivated by this connection, we link our results to stopping power modeling. Although nonlinear stopping formulations exist, they typically assume a near-equilibrium electron background. In our runs, a Maxwellian is driven into a nonequilibrium multiharmonic state. Diffraction narrows the resonant velocity band and modifies trapping/harmonic locking, shifting resonant wave-particle coupling relative to near-equilibrium baselines, with implications for screening and stopping. By extending KEEN physics to the quantum regime, with Wigner-Poisson, we aim to establish a controlled setting to study how quantum diffraction may modify the thresholds, spectra, and saturation of driven sub-$\omega_{pe}$ modes\cite{hu2022kinetic}, with potential implications for dielectric screening and stopping. 

We study a mild-moderate quantum regime where diffraction is not negligible, yet the kinetic phase-space structure (trapping, plateaus, separatrices) remains dynamically accessible\cite{mendoncca2023landau}. This is precisely the regime relevant to surface plasmons in high mobility platforms(e.g., graphene or doped semiconductors)\cite{fang2013gated}, bulk plasmons in warm-dense conditions and laser-driven electron dynamics in solds where collisions are finite but not dominant. In these systems, externally imposed beats or envelopes can create multiharmminc long-lived oscillations that persist after the drive is removed\cite{afeyan2012kinetic}. Our simulations show that, once quantum diffraction is taken into account, the mechanism that stabilizes KEEN-like states changes. The effective resonance narrows, trapping vortices reshape, and harmonic locking shifts, thus changing the conditions under which Landau damping is suppressed\cite{mendoncca2023landau}.

Within this framework, KEEN waves offer an ideal test ground for learning about driven non-linear dynamics of Wigner Poisson in a reduced setting. Classically, a brief drive tuned between the electron-plasma and the electron-acoustic frequencies traps electrons, locks multiple harmonics, and produces long-lived oscillations, which survive thousands of plasma periods after the driver is removed \cite{montgomery2001observation,afeyan2004simulations,johnston2009persistent,afeyan2012kinetic,afeyan2014simulations}. Their multiharmonic structure, rooted in nearly shielded trapped electrons, cannot be captured by fluid models. However, all prior KEEN studies neglect quantum diffraction. Our results show that once the quantum effects are admitted, the very mechanism that stabilizes KEEN waves is re-patterned, diffraction narrows the effective resonance, reshapes trapping vortices, and modifies harmonic locking, thereby shifting the conditions under which Landau damping is suppressed.

By reveling how quantum kinetic physics modulates a paradigmatic nonlinear electrostatic mode, this work advances a unified view of driven, nonequilibrium quantum plasmas. It suggests that there could be opportunities for laser-driven plasmon experiments to seed subplasma responses and provides a framework for translating observed nonlinear structure into predictive transport in warm-dense and solid state platforms.

\section{The non-dimensionalized Wigner-Poisson Model}
We will review in this section the non-dimensionalization of the 1D1V Wigner-Poisson system following \cite{christlieb2025sampling}. This model is used to study quantum mechanical effects on KEEN waves and beyond classical terms, the Wigner-Poisson system features a non-local pseudo-differential operator that encapsulates quantum phenomena such as tunneling and wave-function interference \cite{QuantumPlasmas}. The Wigner-Poisson system can be derived from a quantum many-body approach, as described by kinetic theory molecular dynamics (KTMD) \cite{PhysRevE.90.033104}. The Wigner-Poisson system represents the mean-field or Vlasov approximation in KTMD. Therefore, it is a suitable description of many body quantum systems such as weakly coupled or ideal plasmas with quantum effects. 

Starting with the Wigner-Poisson given in \cite[Chapter 4]{QuantumPlasmas}
\begin{subequations}\label{eq:WP_model_dim}
    \begin{align}
        \frac{\partial \tilde{f}}{\partial \tilde{t}} + \tilde{v} \frac{\partial \tilde{f}}{\partial s} 
&= -\frac{iem_e}{2\pi \hbar^2} \iint d\tilde{v}'\, ds'\, \exp\left( i m_e \frac{(\tilde{v}' - \tilde{v}) s'}{\hbar} \right) \nonumber\\
&\times \left[ \phi\left(s + \frac{s'}{2}\right) - \phi\left(s - \frac{s'}{2}\right) \right] \tilde{f}(s, \tilde{v}', \tilde{t}), \\
\frac{\partial^2 \phi}{\partial s^2} &= -\frac{e}{\epsilon_0} \left( \int d\tilde{v}\, \tilde{f} - n_0 \right), \label{eq:poisson_dim}
    \end{align}
\end{subequations}
Let $\tau$ be a characteristic time scale, $l$ the characteristic length scale, and $\Tilde{\phi}$ the characteristic potential scale. Then, we can define the nondimensional variables and the function as:
\begin{align*}
   \tilde{t} = \tau t,\quad s = l x,\quad \phi = \bar{\phi} \Phi,\quad \tilde{v} = \frac{l}{\tau} v,\quad f = \frac{l}{n_0 \tau} \tilde{f}.
\end{align*}
Substituting them into \eqref{eq:WP_model_dim} yields: 
\begin{subequations}
    \begin{align}
        \frac{\partial f}{\partial t} + v\frac{\partial f}{\partial x} & = \frac{-iC}{2\pi H^2} \int \int dv' dx' \exp{\left(i \frac{v'-v}{H}x'\right)}\nonumber\\
        &\times \left[\Phi \left(x+\frac{x'}{2}\right)+\Phi \left(x - \frac{x'}{2}\right)\right]f(x,v',t) \label{eq:wigner_non} \\
        - \frac{\partial \Phi}{\partial x^2} & = D \left( \int dvf-1 \right) \label{eq:poisson_non}
    \end{align}
\end{subequations}
where the dimensionless parameters are defined as
\begin{align*}
    C = \frac{e\tilde{\phi} \tau^2}{ml^2}, \;\; H = \frac{\tau \hbar}{ml^2}, \;\; D = \frac{enl^2}{\tilde{\phi}\epsilon_0}. 
\end{align*}
Now, to further simplify the system and highlight key physical scalings, we choose the characteristic potential scale, time scale by the plasma frequency $\tau=\omega_{pe}^{-1}$, length scale as the Debye length $l=\lambda_D$, 
\begin{align*}
    \tilde{\phi} = \frac{enl^2}{\epsilon_0}, \, \omega_{pe} = \sqrt{\frac{e^2n}{m_e\epsilon_0}},\, \lambda_D = \sqrt{\frac{\epsilon_0k_BT}{e^2n}}. 
\end{align*}
Such choices make our scaling velocity the thermal velocity and lead to much simpler form of dimensionless parameters: 
\begin{align}\label{eq:non_dim_parameters}
    v = \frac{l}{\tau} = \lambda_D\omega_{pe} = \sqrt{\frac{k_BT}{m_e}} = v_{th}, \, C = \frac{e\tilde{\phi} \tau^2}{ml^2}  = 1, \nonumber \\
    D  = \frac{enl^2}{\tilde{\phi}\epsilon_0} = 1, \, H  = \frac{\hbar}{m_e\lambda^2_D \omega_{pe}}. 
\end{align}
Thus, the non-dimensional Wigner-Poisson system is written as:
\begin{subequations}
    \begin{align}
        \frac{\partial f}{\partial t} + v\frac{\partial f}{\partial x} & = -\frac{i}{2\pi H^2} \int \int dv' dx' \exp{\left(i \frac{v'-v}{H}x'\right)}\nonumber\\
        &\times \left[\Phi \left(x+\frac{x'}{2}\right)+\Phi \left(x - \frac{x'}{2}\right)\right]f(x,v',t) \label{eq:wigner_good} \\
        - \frac{\partial \Phi}{\partial x^2} & =  \int dvf-1  \label{eq:poisson_good}
    \end{align}
\end{subequations}
The right-hand side of \eqref{eq:wigner_good} is a nonlocal term in the form of a double integral, which we refer to as the Wigner potential operator. In implementation, the double integral requires simultaneous access to both variables, making it computationally demanding. This coupling poses a significant challenge for parallelization, as it leads to a high communication cost at each time step.  We note that an approach to address this is adaptive rank solvers, see \cite{christlieb2025sampling}.

\begin{figure}[htbp]
  \centering
        \centering
        \includegraphics[width= \linewidth]{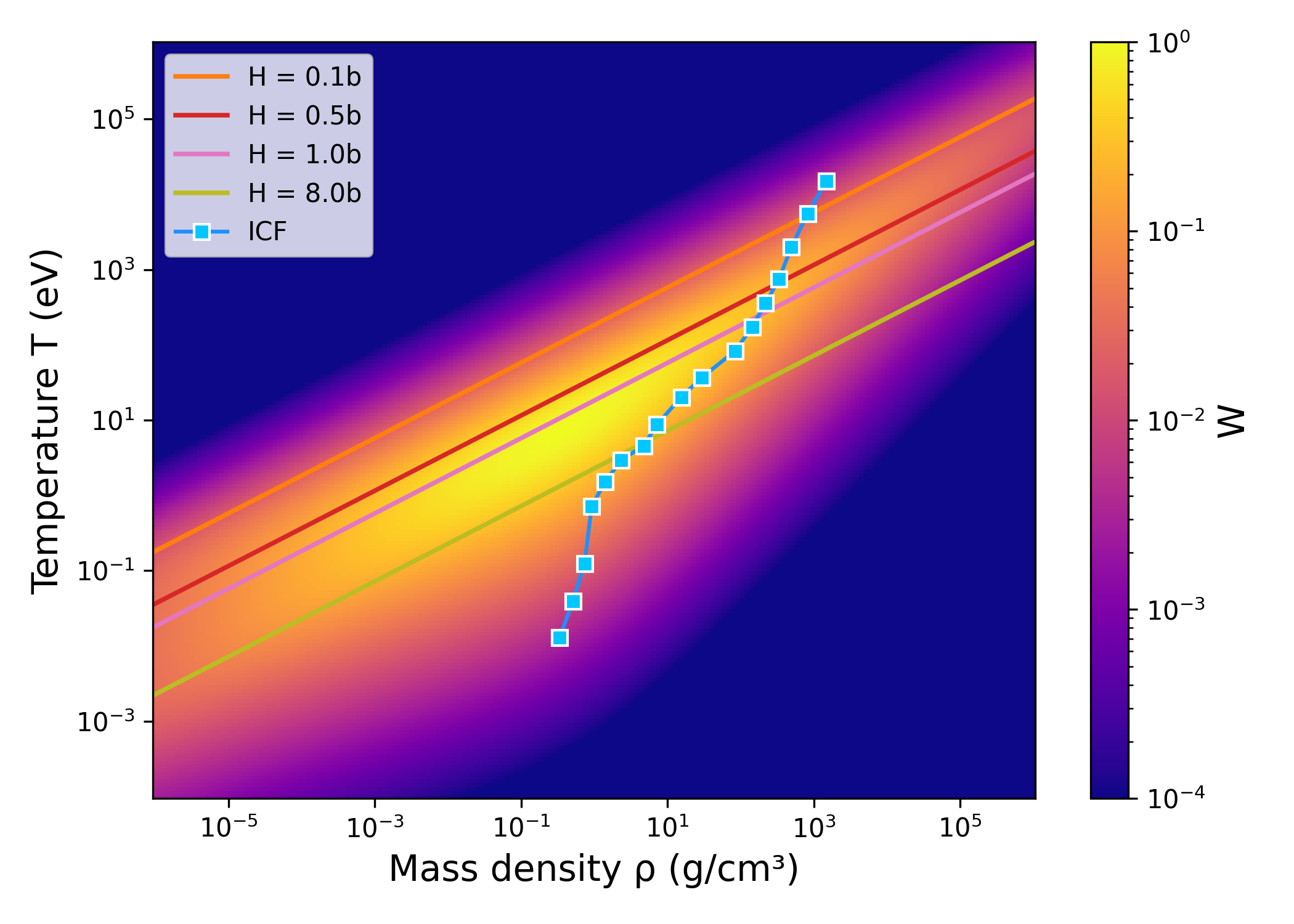}
    \caption{Log-scale heatmap of the Warm-Dense parameter $W = S(\Gamma_{ee})S(\Theta)$, with $S(x)= 2x/(1+x^2)$ for a fully ionized 50-50 deuterium-tritium mixture. Solid lines correspond to four values of the rescaled parameter $ H b$, with $b = \sqrt{m/m_{DT}}$, where $m$ is the reference mass and $m_{DT}$ the mean deuterium-tritium mass. Cyan squares show the ICF data from \cite{haines2024charged}. Axes show mass density $\rho$ (g/$cm^{3}$) and temperature T (eV).}
    \label{fig:hline}
\end{figure}

Note that the dimensionless quantum parameter $H$ defined in \eqref{eq:non_dim_parameters} still needs to be assigned a concrete value. Because our focus is on warm-dense plasmas, we only care about those values of $H$ that occur under the temperature and density conditions relevant to laser-driven electron dynamics. Consequently, we will re-express $H$ explicitly as a function of the plasma temperature and mass density, so that we can identify the numerical range of $H$ appropriate to our warm-dense regime. 
\begin{align*}
    H \; = \; \frac{\hbar}{m_e\lambda^2_D\omega_{pe}} \; = \; \frac{\hbar}{k_BT}\sqrt{\frac{e^2n}{m_e\epsilon_0}}.
\end{align*}
In \cite{murillo2010x}, the author used an expression for $W$ that indicates if the plasma is in the Warm Dense state.  In this work, we express $W$ for a fully ionized 50-50 deuterium-tritium plasma.  Here the mass density $\rho$ is in (g/$cm^{3}$) and temperature T (eV).   The warm-dense parameter $W$ is given by $W = S(\Gamma_{ee})S(\Theta)$, where $\Gamma_{ee}= e^2/(a_e\sqrt{(k_BT)^2+E^2_F}$ is the electron-electron coupling parameter,  $a_e =\left(3/4\pi n_e \right)^{1/3}$ and $\Theta = E_F/k_BT$ and $E_F$ the Fermi energy \cite{bender2021simulation}.   In Figure \ref{fig:hline} we plot the a heat map of the log of the warm-dense parameter $W$, together with four values of $H$ and ICF data points from \cite{haines2024charged}.  One clearly sees that the experimental data for an ICF target passes through the warm-dense state.  The plot also shows that the warm-dense state supports quantum electrons, $H=O(1)$.

\section{Numerical Method}
In this work we solved the Wigner-Poisson system with a second order Strang Splitting firstly proposed in \cite{suh1991numerical}, with theory in \cite{OperatorSplittingWP}.  A full order, as well as a low rank version,  with a structure-preserving Fourier update with high order in space Semi-Lagrangian WENO method was proposed in \cite{christlieb2025sampling}.  In this study, we make use of this full-rank method. For completeness, we briefly review this solver below

The method decouples complex Wigner-Poisson system to several easier sub-problems and solves each sub-problem based on improved solvers. In particular, at each time step, we alternate between pure advection (steps 1 and 4) and a nonlocal update in Fourier space (step 3), with a Poisson solver in between to recompute the electric potential. Let $t^n = n\Delta t$, periodic spatial domain $x \in [a,b]$, symmetric velocity domain $v \in [-L_v, L_v]$. We use a uniform grid. The grid points are defined as $x_i = a + (i-1)\Delta x$ and $v_i = -L_v + (i-1)\Delta v$, with $\Delta x = (b-a)/N_x$ and $\Delta v = 2L_v / (N_v-1)$. 
\begin{enumerate}
    \item Over $\left(t^n, t^n+ \frac{\Delta t}{2} \right)$, we get $f^{(1)}$ by solving the advection equation using a Conservative Semi-Lagrangian WENO scheme proposed in \cite{Qiu-Christlieb}
    \begin{equation}
        \frac{\partial f}{\partial t} + v \frac{\partial f}{\partial x} = 0 \label{eq:adv1}
    \end{equation}
    \item Calculate the density using the solution of the previous step $\int dvf^{(1)}$ and solve the Poisson equation to get $\Phi^{n+1/2}$:
    \begin{equation}
        -\frac{\partial^2 \Phi}{\partial x^2} = \int dv f^{(1)}-1
    \end{equation}
    \item Over $\left( t^n, t^{n+1} \right)$, solve the Wigner potential operator with $f^{(1)}$ and $\Phi^{n+1/2}$ to get $f^{(2)}$
    \begin{eqnarray}
        \frac{\partial f}{\partial t} =&& -\frac{i}{2 \pi H^2}\int \int dv'dx' \exp{\left(i \frac{v'-v}{H}x'\right)}\nonumber\\
        &&\times\left[\Phi\left(x+\frac{x'}{2}\right) - \Phi\left(x-\frac{x'}{2}\right)\right]f(x,v',t) \label{eq:wigterm}
    \end{eqnarray}
    \item Over $\left( t^n + \frac{\Delta t}{2}, t^{n+1} \right)$, we again the advection equation \eqref{eq:adv1} using $f^{(2)}$ to obtain $f^{n+1}$. 
\end{enumerate}
Step 2 is solved using a standard fourth-order finite difference method, with the trapezoidal rule applied to the integration. The methods for the remaining steps are described in the following two subsections.

\subsection{Conservative Semi-Lagrangian}
This section outlines the keypoints of the conservative Semi-Lagrangian (SL) WENO scheme from \cite{Qiu-Christlieb}, which is used to solve the advection equation \eqref{eq:adv1}.

The main idea of the SL framework originates from the method of characteristics, where the advection equation is solved by tracing characteristics backward in time, governed by the equation:
\[
\frac{dx}{dt} = v, 
\]
since the solution of the advection equation remains constant along these characteristic curves.
Let $f^n_j(x)$ denote the semi-discrete solution at time $t^n, \, v = v_j$ and $f^{n}_{i,j} = f^n_j(x_i)$. For any fixed $v_{j}$, to compute $f^{(1)}_{i,j}$, one traces back along the characteristic to the departure point $x_d := x_i - v_j (\frac{\Delta t}{2})$ at time $t^{n} + \frac{\Delta t}{2}$ with 
$f^{(1)}_{i,j} = f^n_j(x_d)$. The value of $f^n_j(x_d)$ will then be evaluated by some interpolations and in our case, we use specific the conservative WENO5 form; see more details in \cite{Qiu-Christlieb} and \cite{christlieb2025sampling}. 

\subsection{Fourier Update}
In this section, we address Step 3. To overcome the computational challenge posed by the nonlocal Wigner potential operator, we take the Fourier transform, which yields an ordinary differential equation (ODE) that admits an analytic solution in the discrete time setting. Further implementation details can be found in \cite{christlieb2025sampling}.

Starting from the Fourier transform with respect to the velocity space $v$ of \eqref{eq:wigterm}, we obtain:

\begin{eqnarray}
    \frac{\partial \Tilde{f}}{\partial t} =&&  -\frac{i}{H^2} \int dx' \delta\left(-k_v - \frac{x'}{H} \right)\nonumber\\
    &&\times \left[\Phi\left(x + \frac{x'}{2}\right) - \Phi\left(x - \frac{x'}{2}\right)\right] \tilde{f}(x, k_v, t). \label{eq:wigtermfourier}
\end{eqnarray}
With the change of variable $\hat{x} = -k_v - \frac{x'}{H}$ and $d\hat{x} = -\frac{1}{H}dx'$ we get
\begin{equation}
    \frac{\partial \tilde{f}}{\partial t} = \frac{i}{H} \left[\Phi\left(x + \frac{Hk_v}{2}\right) - \Phi\left(x - \frac{H k_v}{2}\right)\right] \tilde{f}(x, k_v, t).
\end{equation}
Let 
\[
g(x,k_v,H) := \frac{i}{H}  \left[\Phi\left(x + \frac{H k_v}{2}\right) - \Phi\left(x - \frac{H k_v}{2}\right)\right].
\]
Using the frozen field approximation, which treats our potential `frozen' over one time step, we can get the following closed-form expression for the update
\begin{eqnarray}\label{eq:fourier_update}
   \tilde{f}(x, k_v, t + \Delta t) = \tilde{f}(x, k_v, t) \exp \left ( g(x,k_v,H)  \Delta t \right ).
\end{eqnarray}
In Step 3,  let $\tilde{f}(x, k_v, t)$ denote the Fourier transform of $f^{(1)}$ with respect to the velocity. We replace $\Delta t$ by $\frac{\Delta t}{2}$, and then obtain $f^{(2)}$ by taking the inverse Fourier transform of the corresponding update. Note that the right-hand-side of \eqref{eq:fourier_update} is conjugate symmetric, since both the Fourier transform of a real-valued function and the exponential term $\exp \left ( g(x,k_v,H)  \Delta t \right )$ are conjugate symmetric, i.e.  
\begin{align*}
   \tilde{f}(x, k_v, t) & = \overline{\tilde{f}(x, -k_v, t)}, \\
\exp \left ( g(x,k_v,H)  \Delta t \right ) &= \overline{\exp \left ( g(x,-k_v,H)  \Delta t \right )}. 
\end{align*}
This symmetry implies that the inverse Fourier transform of $\tilde{f}(x, k_v, t + \Delta t)$ yields a real-valued function. This structure is preserved with our improved Fourier update proposed in \cite{christlieb2025sampling} by shifting the order of frequencies and discarding the Nyquist(highest) frequency; see more implementation details in \cite{christlieb2025sampling}.

\section{Results}

To model KEEN waves, we introduce an external potential into the Wigner-Poisson system. The resulting equation becomes:

\begin{subequations}
   \begin{align}
    \frac{\partial f}{\partial t} + v \frac{\partial f}{\partial x} &= \Theta[\Phi_{sc} + \Phi_{ext}] f  \label{eq:wigner} \\
    - \frac{\partial^2 \Phi_{sc} }{\partial x^2} & = \int dv  f - 1 \label{eq:poisson}
\end{align} 
\end{subequations}
with
\begin{eqnarray}
        \Theta[\Phi]f(x,v,t) =&& -\frac{i }{2 \pi H^2} \int \int dv' dx' \exp\left( i \frac{v' - v}{H} x' \right) \nonumber\\
        &&\times \left[ \Phi\left(x + \frac{x'}{2}\right) - \Phi\left(x - \frac{x'}{2}\right)\right]\nonumber\\
        &&\times f(x,v', t)
\end{eqnarray}

In the Wigner-Poisson system, $f_w$ is the Wigner distribution function, $\Phi_{sc}$ is the self-consistent potential, $\Phi_{ext}$ is the external potential, and $H$ is a non-dimensionalized $\hbar$. Wigner describes the evolution of a particle distribution function; it can take negative values.

For the external potential to drive a KEEN wave, we will use the one used in \cite{Yingda} given by

\begin{equation}
    \Phi_{ext} = \frac{A_d(t)}{k}\cos(kx-\omega t)  \label{eq:extphi}
\end{equation}

where 
\begin{equation}
         A_d(t)=
 \begin{cases} 
      0.4\frac{1}{1+\exp{-40(t-10)}} & 0 < t < 60 \\
      0.4\left( 1-\frac{1}{1+\exp{-40(t-110)}} \right) & 60\leq t 
   \end{cases}
\end{equation}

The initial condition is a spatially uniform Maxwellian 
\begin{equation}
    f_0(x,v) = \frac{1}{\sqrt{2\pi}}\exp(-\frac{v^2}{2}), \;\; (x,v ) \in [0, 8\pi] \times [-8, 8]
\end{equation}

All diagnostics were extracted from a single run in the uniform phase-space domain $(x,v) \in [0, 8\pi] \times [-8, 8]$. The space and velocity were discretized with $N_x = N_v = 4096$ mesh, giving $\Delta x = \frac{8\pi}{2048}$ and $\Delta v = \frac{16}{2048}$, with a CFL of 50. The simulation started at $t=0$ and ended in $t=600$

\subsection{Classical Limit}

\begin{figure}[htbp]
  \centering
        \centering
        \includegraphics[width=0.8\linewidth ]{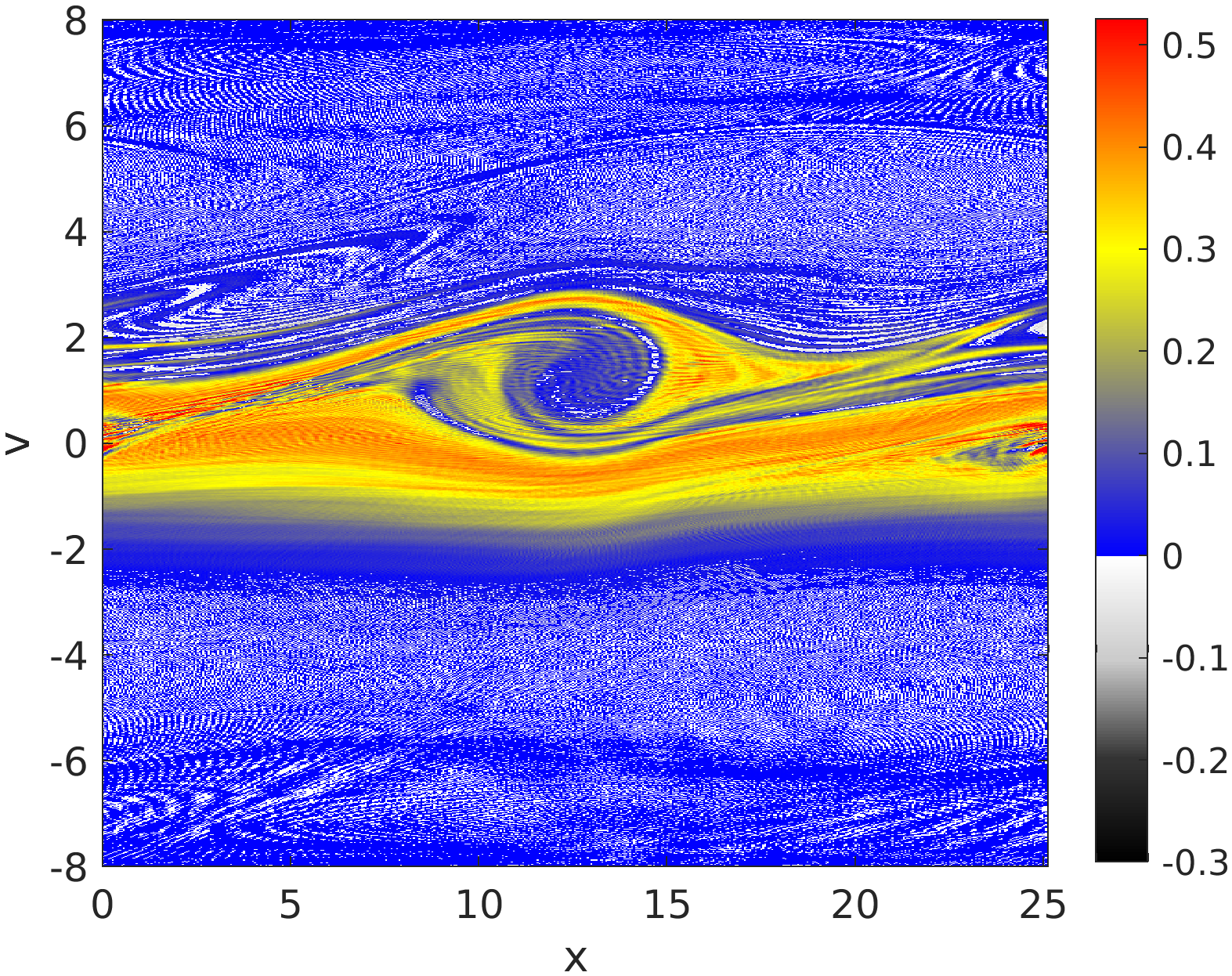}
    \caption{Phase-spaces of KEEN wave at t=60, showing that for shorter times, $H=0.1$ with $Nx=Nv=4096$ simulates fairly well Vlasov-Poisson as in \cite{Yingda}}
    \label{fig:classical_limit}
\end{figure}

In order to verify that the code is reproducing reasonable results, we validate it with the results from \cite{suh1991numerical} and \cite{Yingda}. Comparing phase-space snapshots in \cite{Yingda} we can see that for lower times with $H=0.1$ at $Nx=4096$ is resolved, shown in Fig \ref{fig:classical_limit}

\subsection{Refinement study}

\begin{figure*}[htbp]
    \centering
    \begin{subfigure}{0.45\textwidth}
        \centering
        \includegraphics[width=\linewidth]{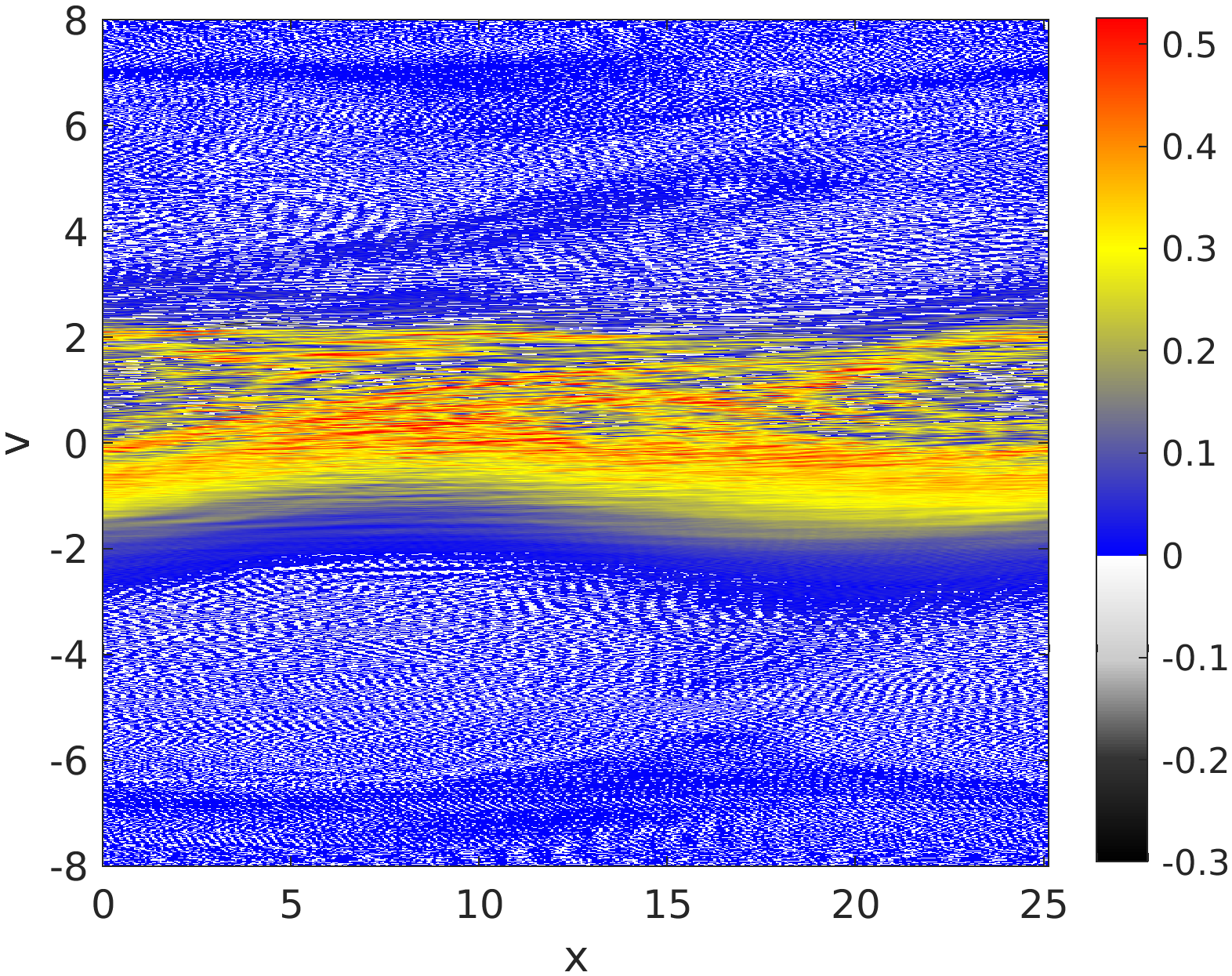}
        \caption{$H = 0.5$, and $Nx = 1024$.}
        \label{fig:refinement_05_1024}
    \end{subfigure}
    \begin{subfigure}{0.45\textwidth}
        \centering
        \includegraphics[width=\linewidth]{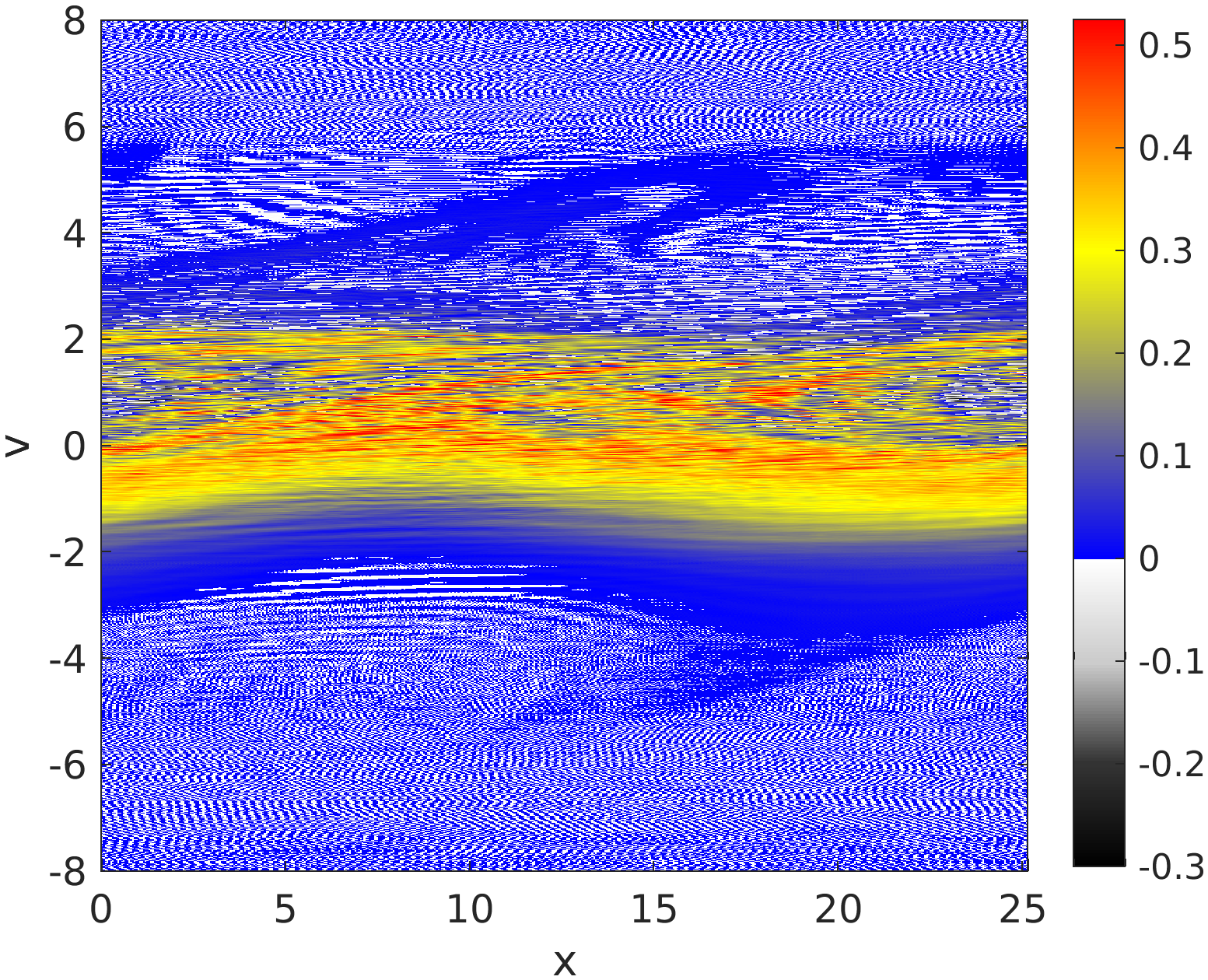}
        \caption{$H = 0.5$, and $Nx =4096$.}
        \label{fig:refinement_05_4096}
    \end{subfigure}
    
    \begin{subfigure}{0.45\textwidth}
        \centering
        \includegraphics[width=\linewidth]{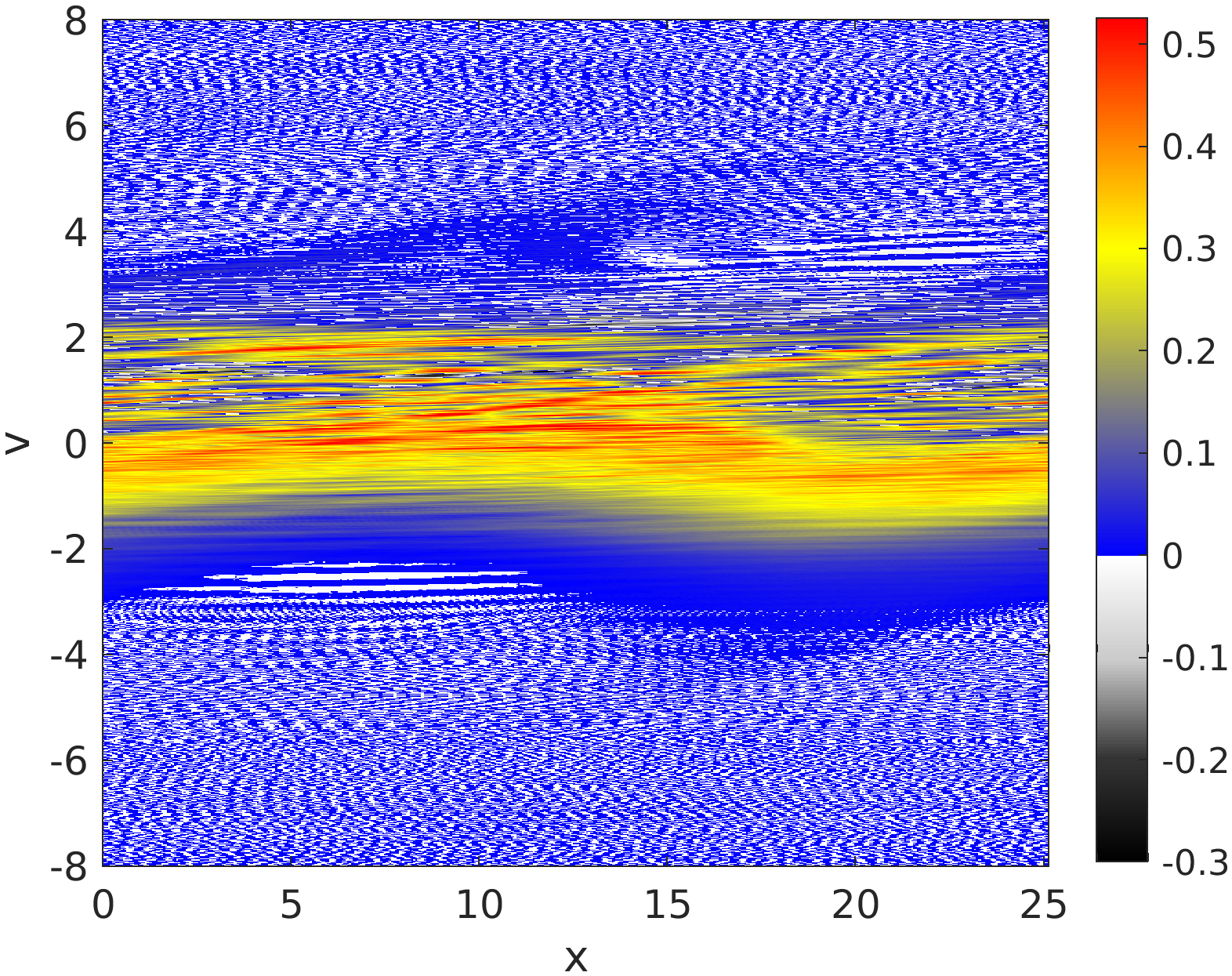}
        \caption{$H = 1$, and $Nx = 1024$.}
        \label{fig:refinement_1_1024}
    \end{subfigure}
    \begin{subfigure}{0.45\textwidth}
        \centering
        \includegraphics[width=\linewidth]{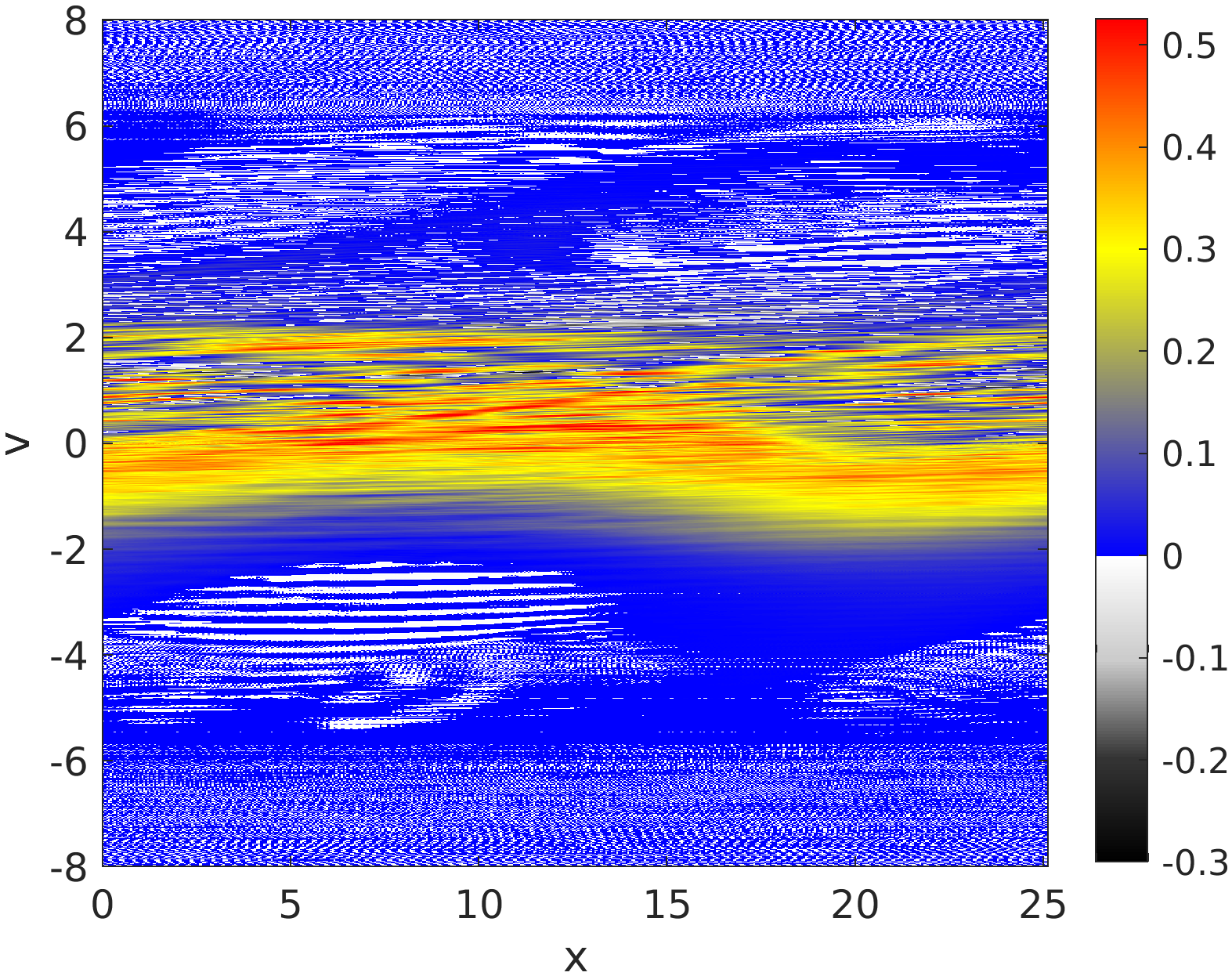}
        \caption{$H = 1$, and $Nx =4096$.}
        \label{fig:refinement_1_4096}
    \end{subfigure}  

    \begin{subfigure}{0.45\textwidth}
        \centering
        \includegraphics[width=\linewidth]{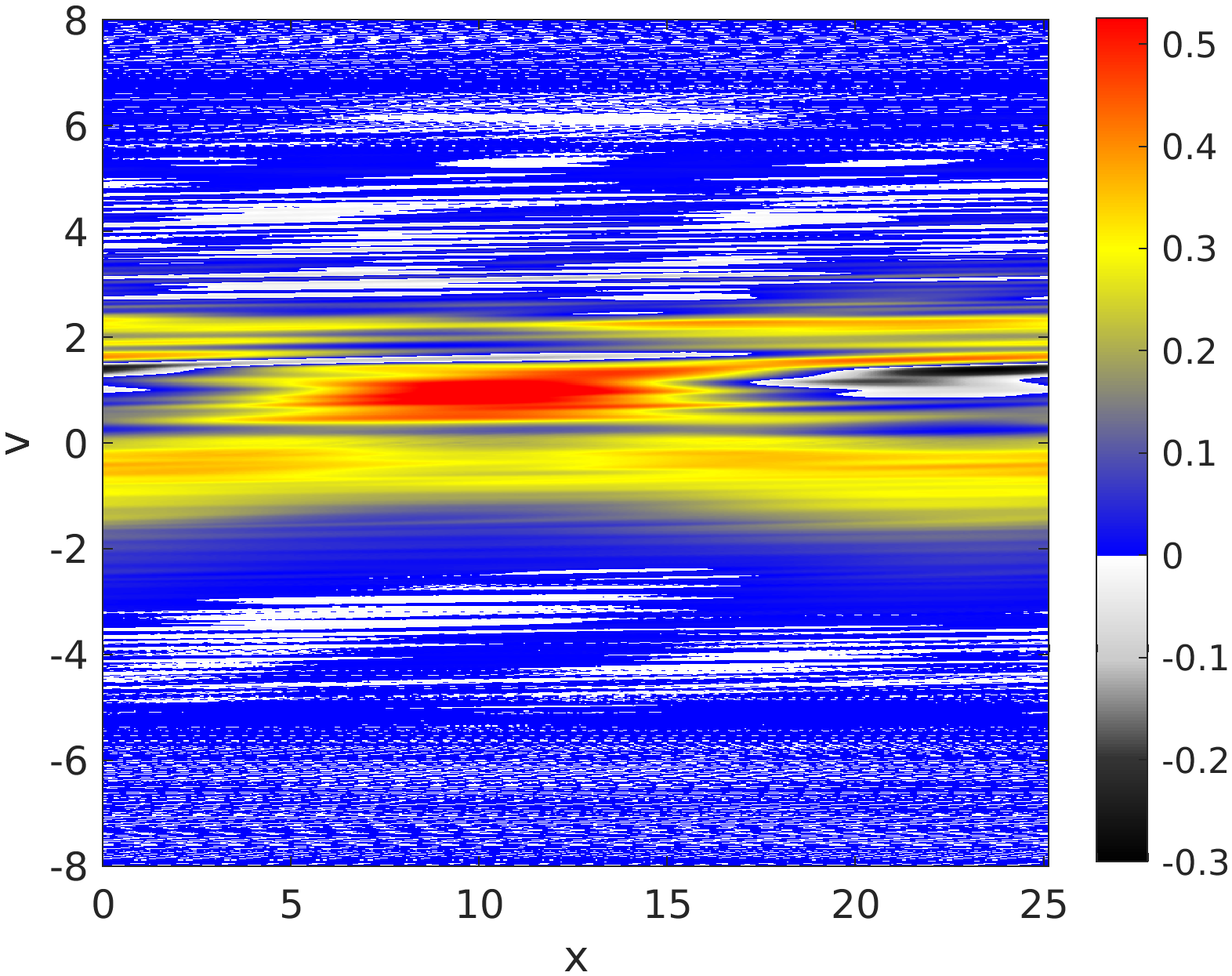}
        \caption{$H = 8$, and $Nx = 1024$.}
        \label{fig:refinement_8_1024}
    \end{subfigure}
    \begin{subfigure}{0.45\textwidth}
        \centering
        \includegraphics[width=\linewidth]{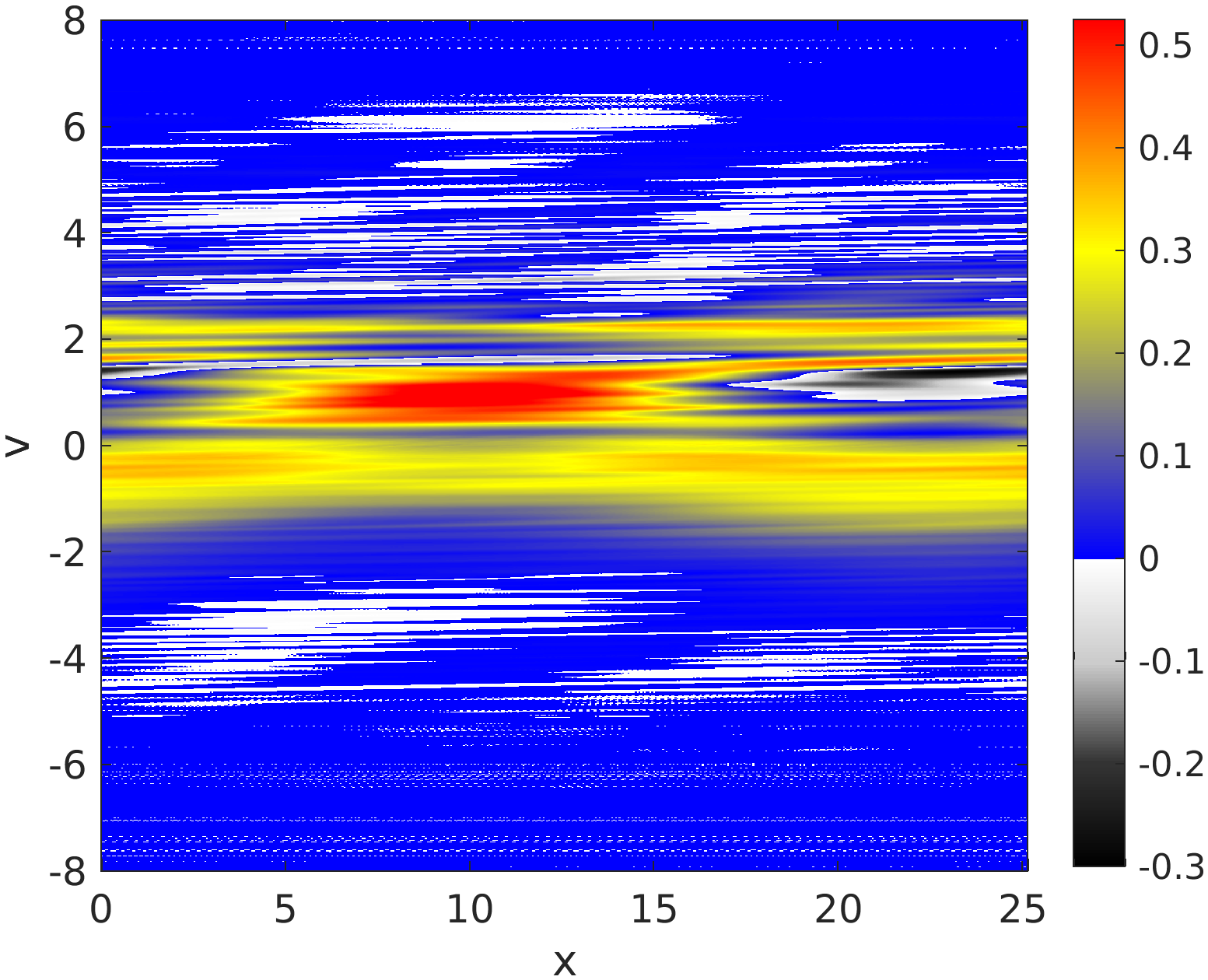}
        \caption{$H = 8$, and $Nx =4096$.}
        \label{fig:refinement_8_4096}
    \end{subfigure}
    \caption{Phase-spaces solutions of the Wigner-Poisson system for KEEN waves at t=120. For $H\approx1$ we see that we need a a finer mesh for the system is fully resolved, but for $H=8$ we see that the system is resolved at within a mesh size of 2048. What we see is that at 2048 the baseline physics are resolved, whereas the fine interference fringes at large $|v|$ and the sharp edges of the separatrix sharpen progressively with resolution. For the $H=0.5$ case, a finer mesh is needed. }
\end{figure*}

To verify our numerical conclusions, we repeated the snapshot $t=120$ on two spatial-velocity grids, $N_x=N_v=1024$ and $4096$, for the three quantum parameters that surround the regime of interest ($H=0.5,1,8$).  The resulting phase-space densities are shown in Figs.~\ref{fig:refinement_05_1024}--\ref{fig:refinement_8_4096}.  We note that unlike Vlasov, Wigner appears to have a smallest length scale established by $H$.  This implies the results can be converged in time and space.  This was first observed in \cite{christlieb2025sampling}.

The central yellow band is captured equally well on both grids.  What changes with resolution are the fine interference fringes in the high--velocity tails and the sharpness of the separatrixes. As expected, the smaller the quantum parameter, the more rapidly those fringes oscillate and the more mesh points are needed to resolve them.  In particular, at $H=0.5$ the coarser grid washes out a noticeable amount of structure in both the upper and lower velocity lobes; a mesh of $4096$ points is required for the small-scale quantum ripples to converge. At $H=1$ the island core and the broad pedestal around it have already converged at $1024$, while the finer grid only sharpens peripheral oscillations that have negligible impact on integrated quantities such as electrostatic energy, and at $H=8$ the distribution is smooth enough that the two grids are quantitatively indistinguishable. We therefore adopt $N_x\!=\!N_v\!=2048$ for the production runs reported in the main text: this resolution captures all physically relevant features for \(H\ge1\) and provides a conservative compromise between accuracy and computational cost for the more demanding case $H=0.5$.

\subsection{Energetics and stationarity}

\begin{figure*}
    \centering
    \includegraphics[width=0.7\linewidth]{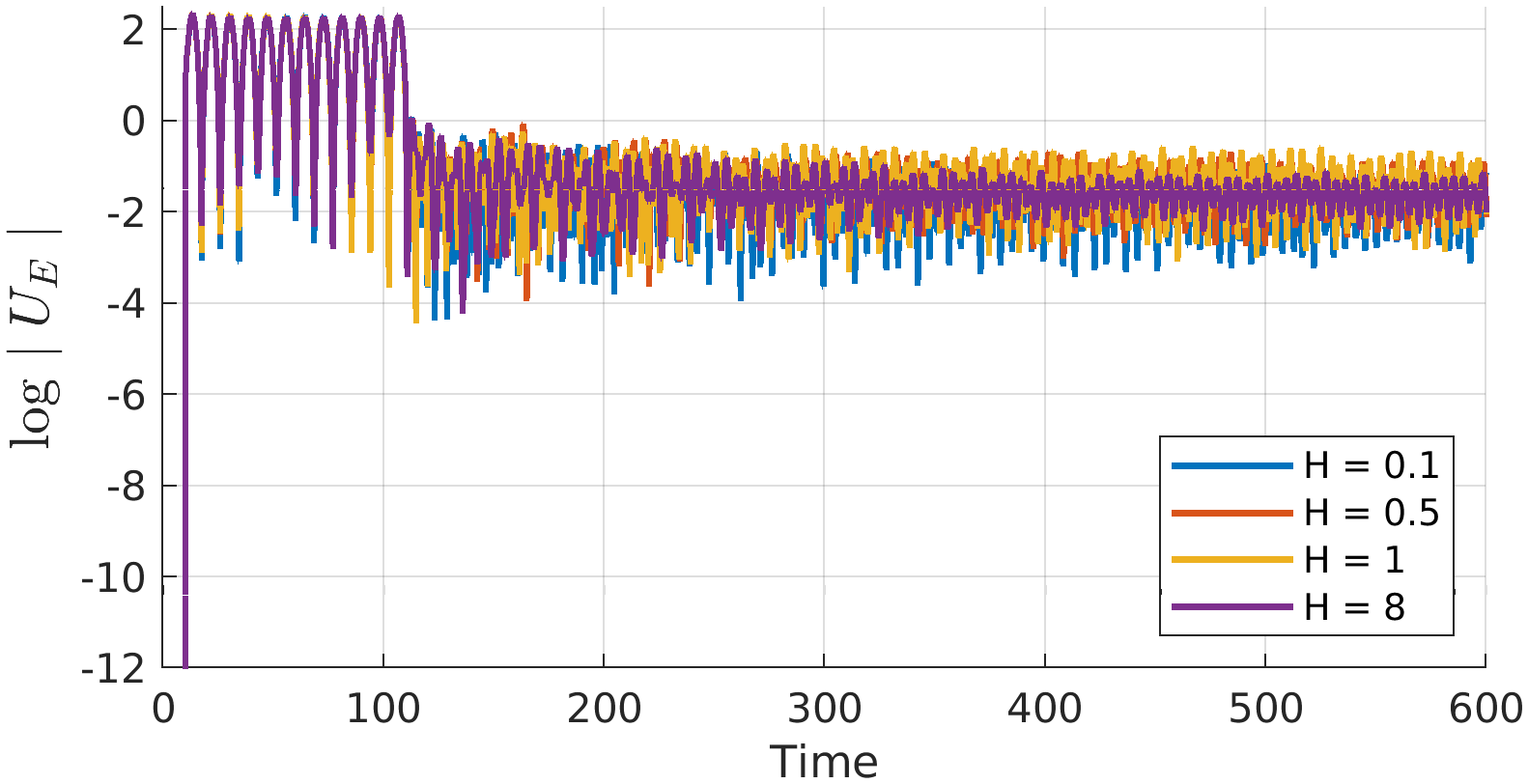}
    \caption{Electrostatic energy for different values of H}
    \label{fig:U_all_H_2048}
\end{figure*}

\begin{figure*}[htbp]
    \centering
    \begin{subfigure}{0.8\textwidth}
        \centering
        \includegraphics[width=\linewidth]{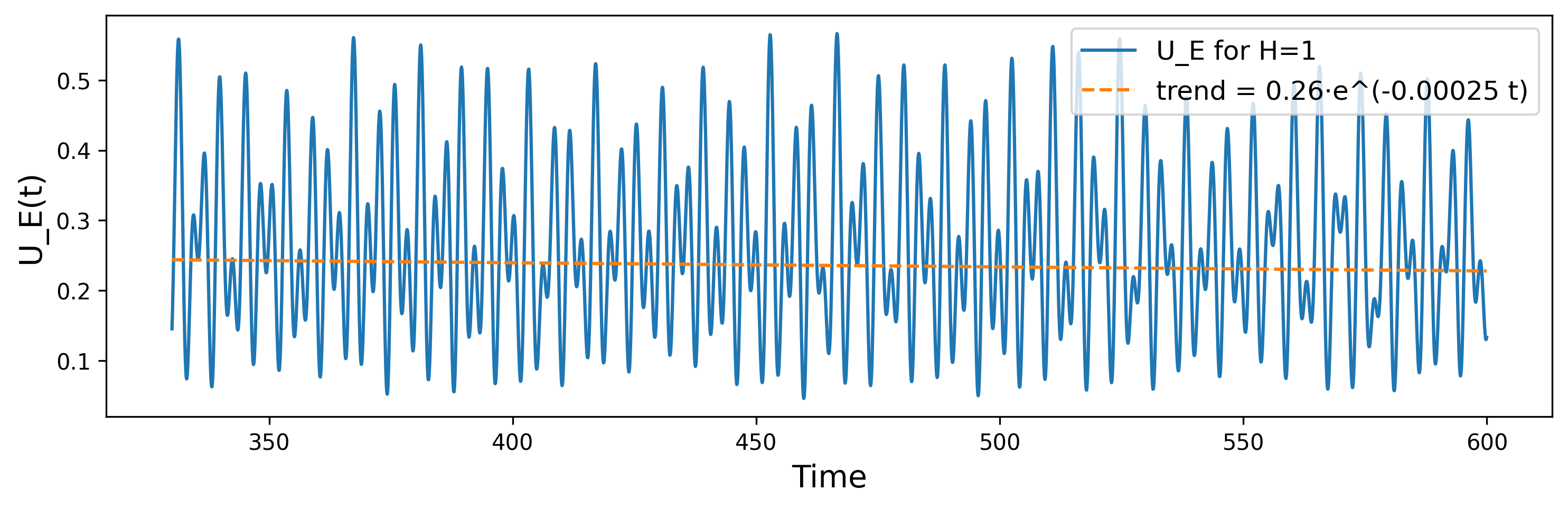}
        \caption{$H = 1$}
        \label{fig:stationaryH1}
    \end{subfigure}

    \begin{subfigure}{0.8\textwidth}
        \centering
        \includegraphics[width=\linewidth]{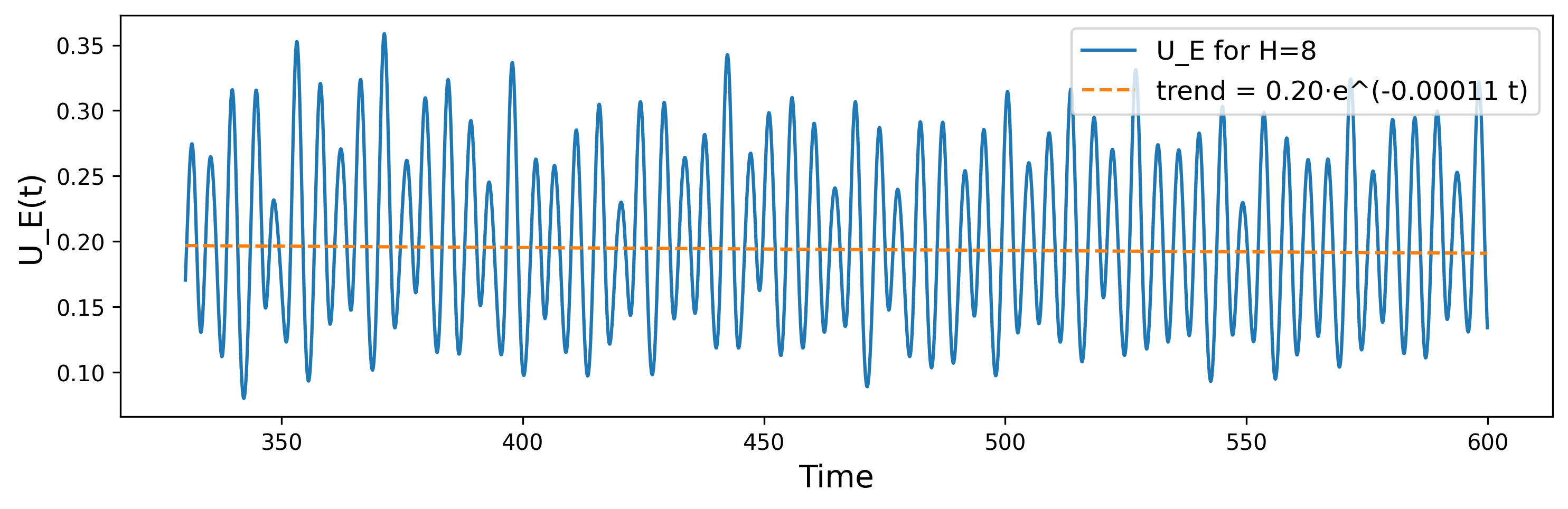}
        \caption{$H = 8$}
        \label{fig:stationaryH8}
    \end{subfigure}
    \caption{ For $t>330$ the electrostatic energy reaches a stationary state, the trapped particles and the self consistent potential are no longer evolving.}
\end{figure*}

Figure~\ref{fig:U_all_H_2048} tracks the electrostatic energy
$U_{E}(t)=\tfrac12\int E^{2}\,dx$ on a logarithmic scale for four values of $H$.  During the drive ($t<60$) all cases rise in unison, confirming that the ponderomotive pump deposits the same energy regardless of quantum strength.  Once the drive is removed, two clear $H$ dependent trends appear: (i) the peak-to-trough amplitude of the subsequent oscillations shrinks with
$H$, and (ii) the envelope of those oscillations damps out faster. Looking over the exponential fit in Figures~\ref{fig:stationaryH1} and \ref{fig:stationaryH8} confirms that reaches a stable, time-independent level by about $t \approx 330$; beyond that point the system remains in its long-time quasi-stationary state.
In \cite{afeyan2012kinetic} the post-drive oscillations are sustained by trapped particles. Taking into account quantum effects, these particles acquire an additional phase-mixing channel via diffraction and tunneling. The larger $H$ therefore accelerates the detrapping process and drives the system to a lower residual energy level.

\subsection{Remnant Harmonics of the Drive as a function of H}

The Fourier modes of the electric field is given by

\begin{equation}
    \resizebox{\columnwidth}{!}{$
  log_nFM(t)= \log_{10}\!\Bigl(\tfrac{1}{L}\sqrt{|\!\int_0^L E\sin(knx)\,dx|^2
      +|\!\int_0^L E\cos(knx)\,dx|^2}\Bigr)$}
\end{equation}

\begin{figure*}[htbp]
  \centering
    \begin{subfigure}{0.45\textwidth}
        \centering
        \includegraphics[width=\linewidth]{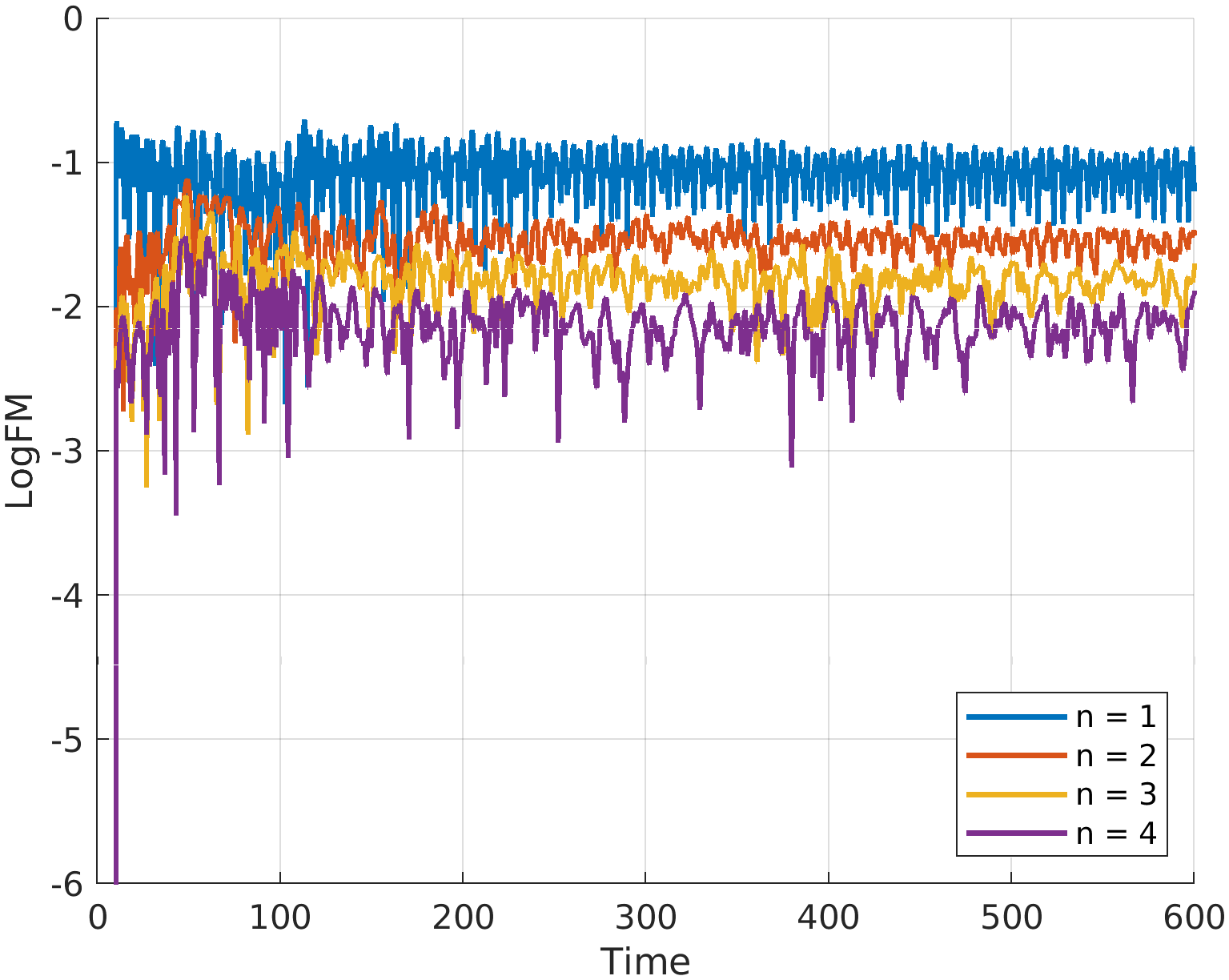}
        \caption{First 4 Fourier modes for $H = 0.5$}
        \label{fig:FM_H=05}
    \end{subfigure}
    \begin{subfigure}{0.45\textwidth}
        \centering
        \includegraphics[width=\linewidth]{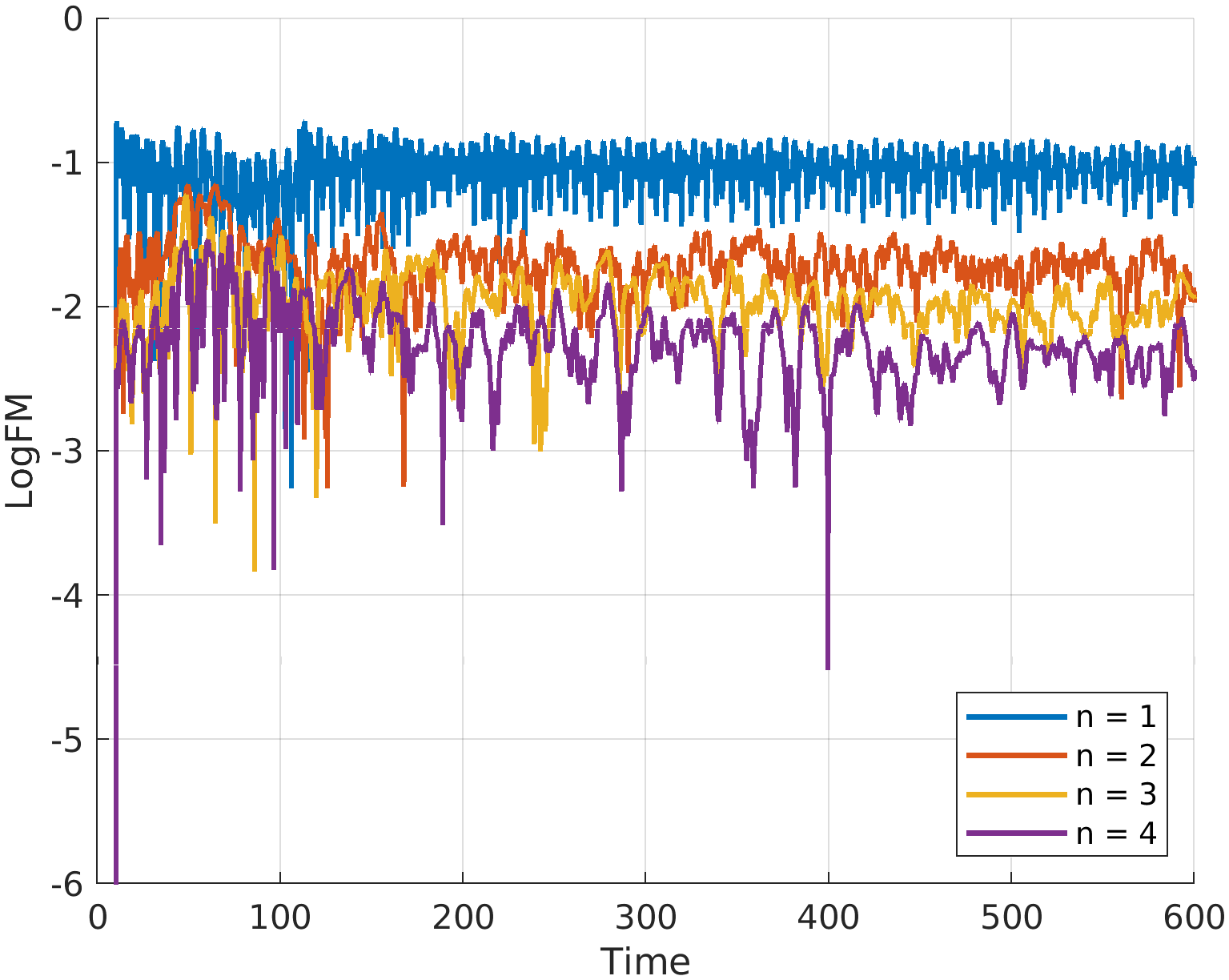}
        \caption{First 4 Fourier modes for $H = 1$}
        \label{fig:FM_H=1}
    \end{subfigure}
    \begin{subfigure}{0.45\textwidth}
        \centering
        \includegraphics[width=\linewidth]{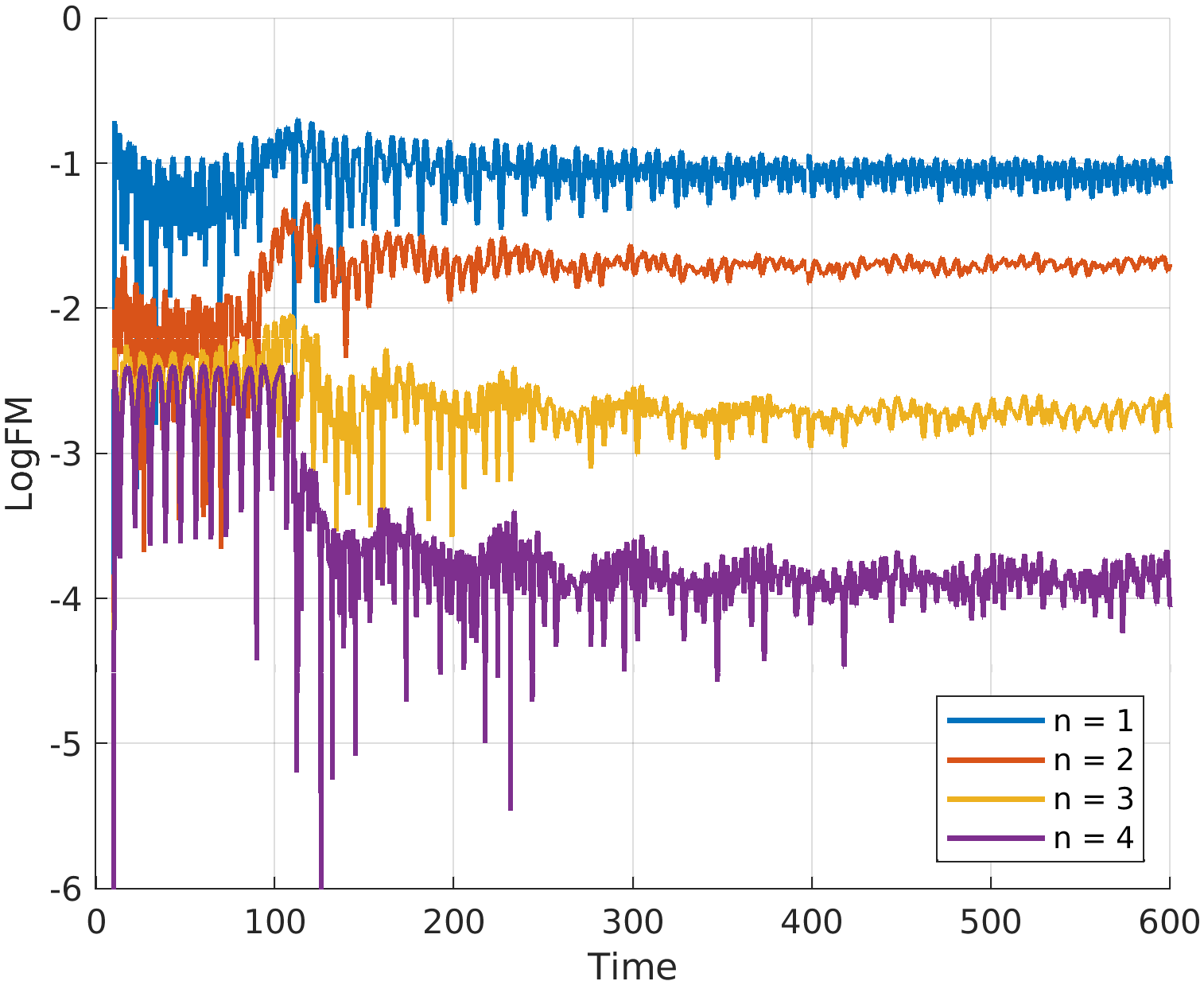}
        \caption{First 4 Fourier modes for $H = 8$}
        \label{fig:FM_H=8}
    \end{subfigure}
    \caption{The first four Fourier modes of the electric field for KEEN waves in Wigner-Poisson are shown above for $H=0.5$, $H=1$, and $H=8$. The strength of the third and forth modes are weaker for $H=8$ than for $H=0.5$ which behaves like Vlasov but they still persist for long time.}
\end{figure*}

In (Figs.~ \ref{fig:FM_H=05}--\ref{fig:FM_H=8}) tell us how much power sits in each harmonic of the electric field.  For $H=0.5$ the first four harmonics sit within roughly one order of magnitude of each other and change little with time, which is what we expect for a strong, long‑lived KEEN wave.  When $H$ increases to 1, the third and fourth harmonics drop. And when $H$ increases to 8, it knocks even the second harmonic down, leaving the driver fundamental as the only significant component. The quantum effects damps small-scale structures in phase-space. Because higher harmonics correspond to finer spatial ripples, they are the first to suffer, leaving only the driver fundamental when $H$ is large enough.

\subsection{A Wavelet time series analysis}
\begin{figure*}[htbp]
    \centering
    \begin{subfigure}{0.8\textwidth}
        \centering
        \includegraphics[width=\linewidth]{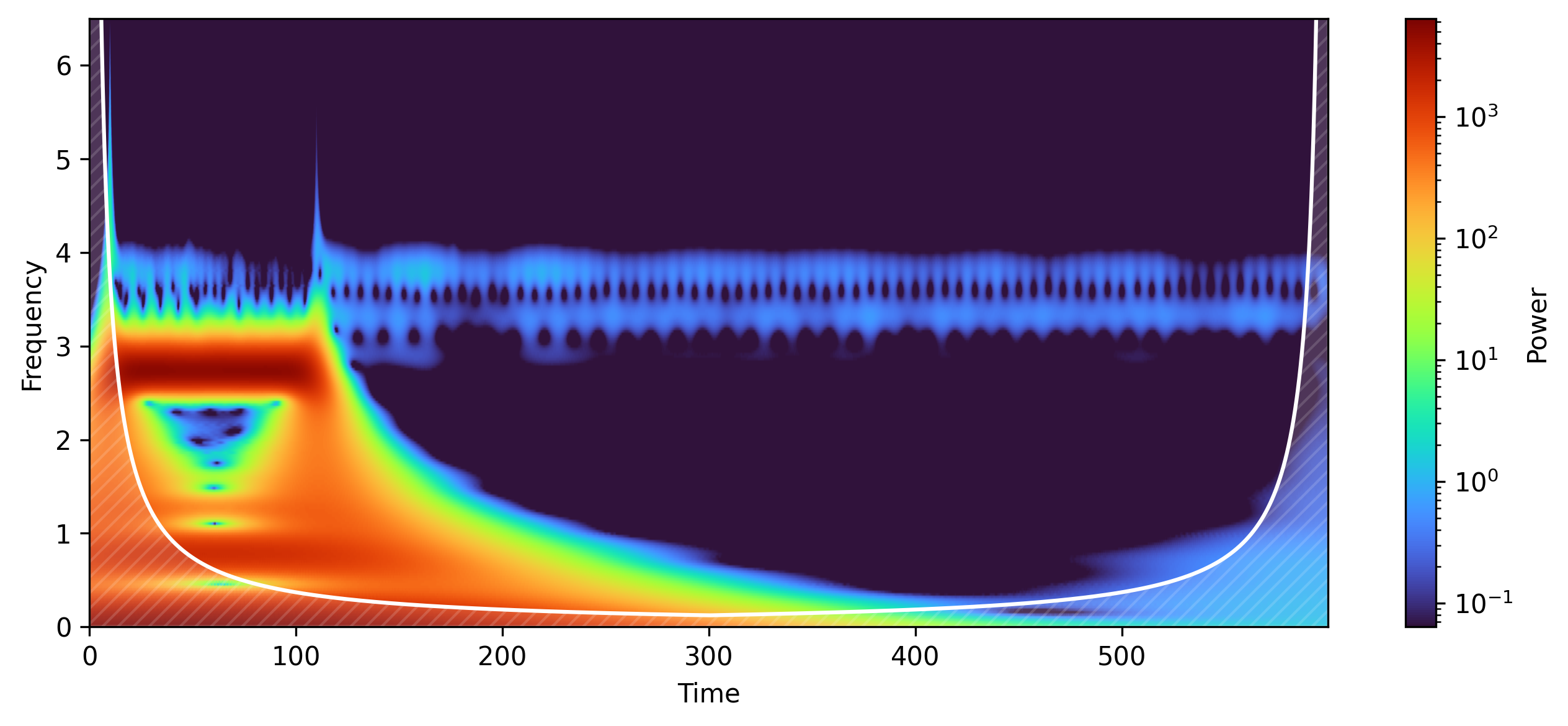}
        \caption{$H = 1$}
        \label{fig:wavelet_H=1}
    \end{subfigure}

    \begin{subfigure}{0.8\textwidth}
        \centering
        \includegraphics[width=\linewidth]{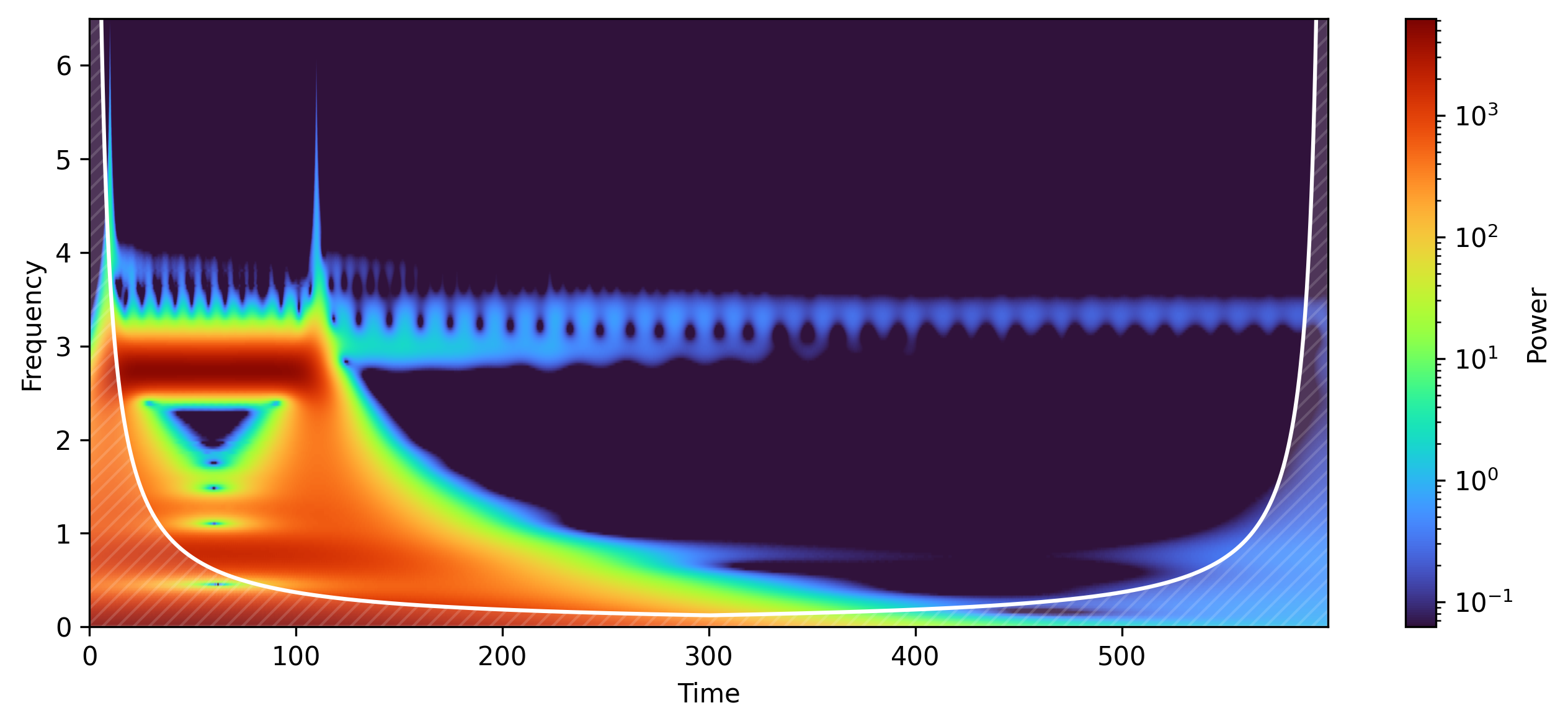}
        \caption{$H = 8$}
        \label{fig:wavelet_H=8}
    \end{subfigure}
    
    \caption{ Wavelet spectrum of electrostatic energy, for the case H=1, its shown a persistent trapped particle vortex. For the case H=8 diffraction and tunneling quickly smears vortexes, higher harmonics vanish within t~300, leaving only a weak, oscillation. Thus larger H accelerates damping of nonlinear KEEN structure}
\end{figure*}

Using a Continuous Wavelet Transform for (Figs.~\ref{fig:wavelet_H=1}--\ref{fig:wavelet_H=8})
\begin{subequations}
   \begin{align}
    W_u(s,\tau) &= \frac{1}{\sqrt{|s|}}\int^\infty_{-\infty}U_E(t)\:\psi^*\left(\frac{t-\tau}{s}\right)dt \label{eq:wavelet}\\
    \psi(t) &= \pi^{-\frac{1}{4}}\exp\left(i\omega_0 t\right) \exp\left(-\frac{t^2}{2\sigma^2}\right) \label{eq:psi}
    \end{align} 
\end{subequations}
condense the foregoing observations into a single diagnostic.  For $H=1$, power fills a triangular region bounded above by the driver frequency. At $H=8$ this broadband patch contracts into a single bright ridge centered on the drive, visual confirmation that only one coherent mode survives.

\subsection{Plasma Density for different quantum intensities}

\begin{figure*}[htbp]
    \centering
    \begin{subfigure}{0.40\textwidth}
        \centering
        \includegraphics[width=\linewidth]{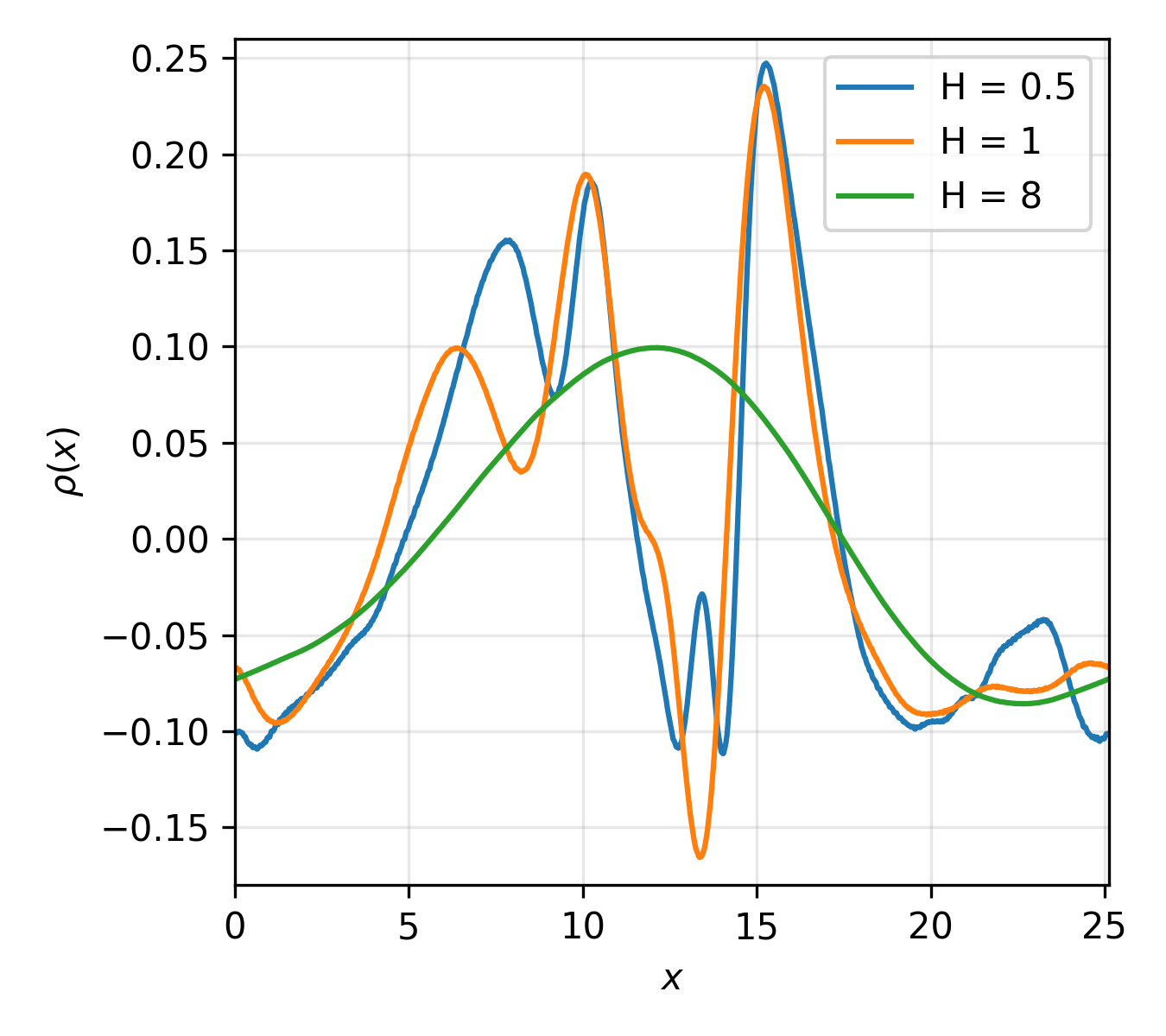}
        \caption{$t=60$.}
        \label{fig:rho60}
    \end{subfigure}
    \begin{subfigure}{0.40\textwidth}
        \centering
        \includegraphics[width=\linewidth]{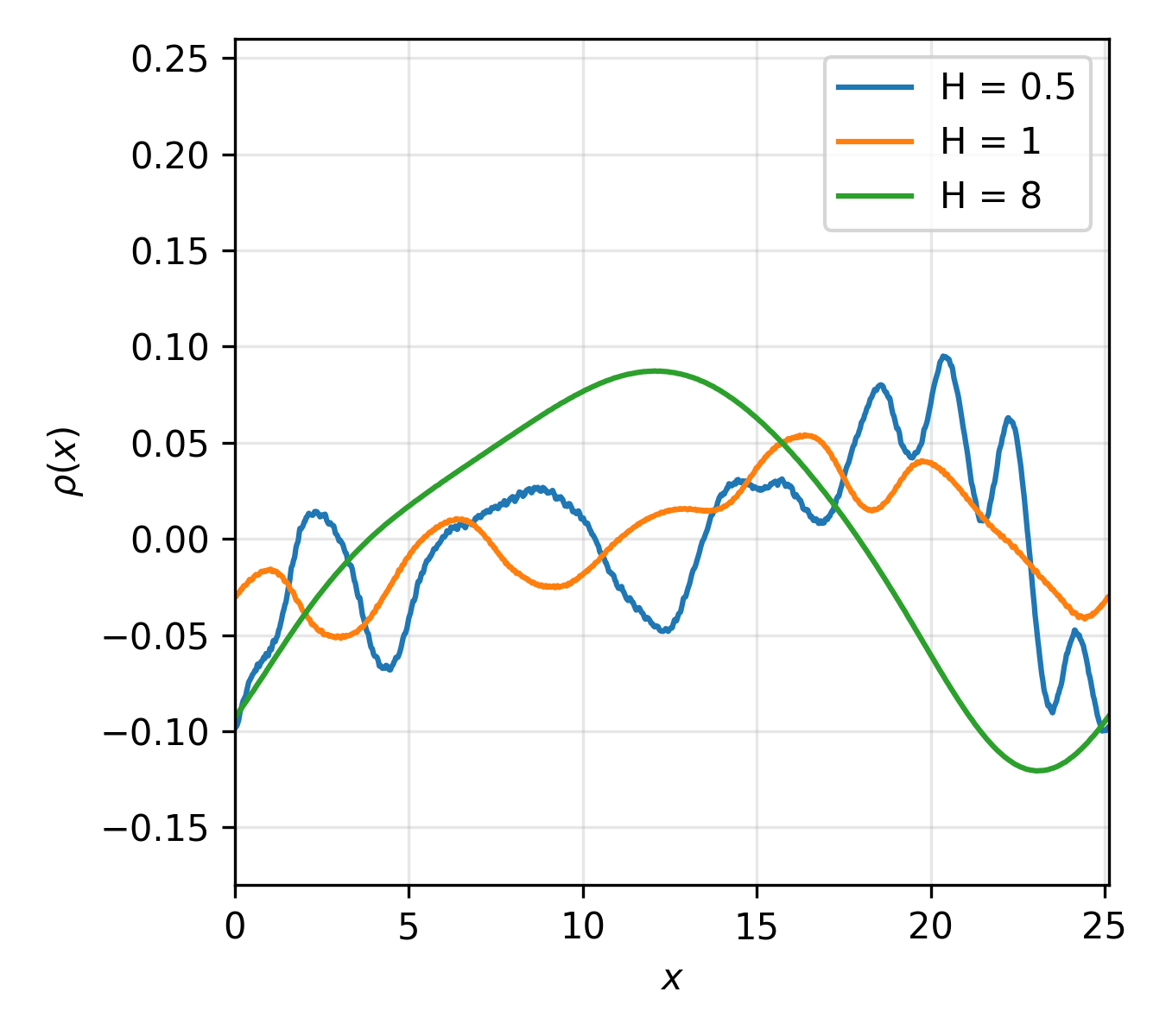}
        \caption{$t=120$.}
        \label{fig:rho120}
    \end{subfigure}
    
    \begin{subfigure}{0.40\textwidth}
        \centering
        \includegraphics[width=\linewidth]{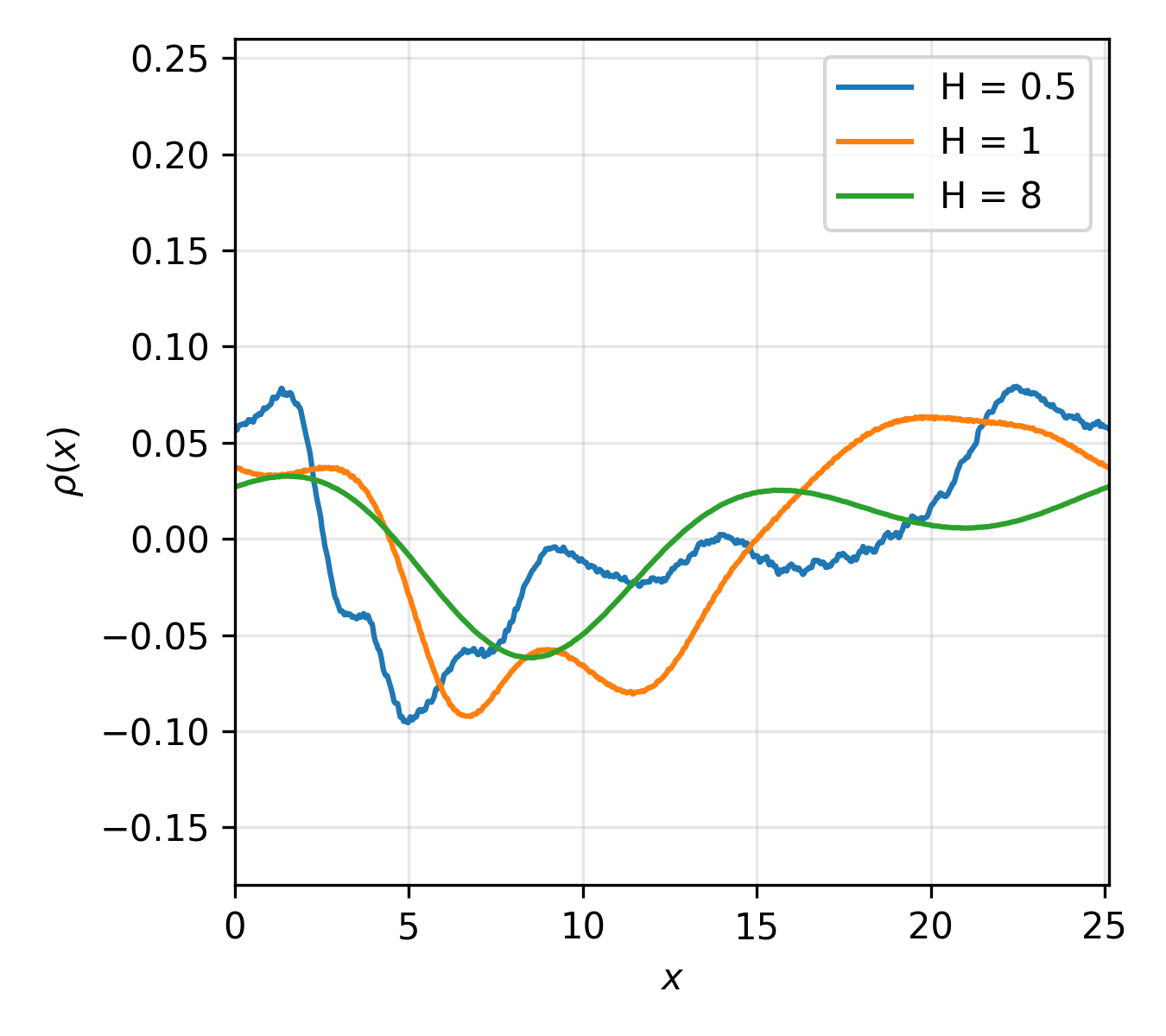}
        \caption{$t=240$.}
        \label{fig:rho240}
    \end{subfigure}
    \begin{subfigure}{0.40\textwidth}
        \centering
        \includegraphics[width=\linewidth]{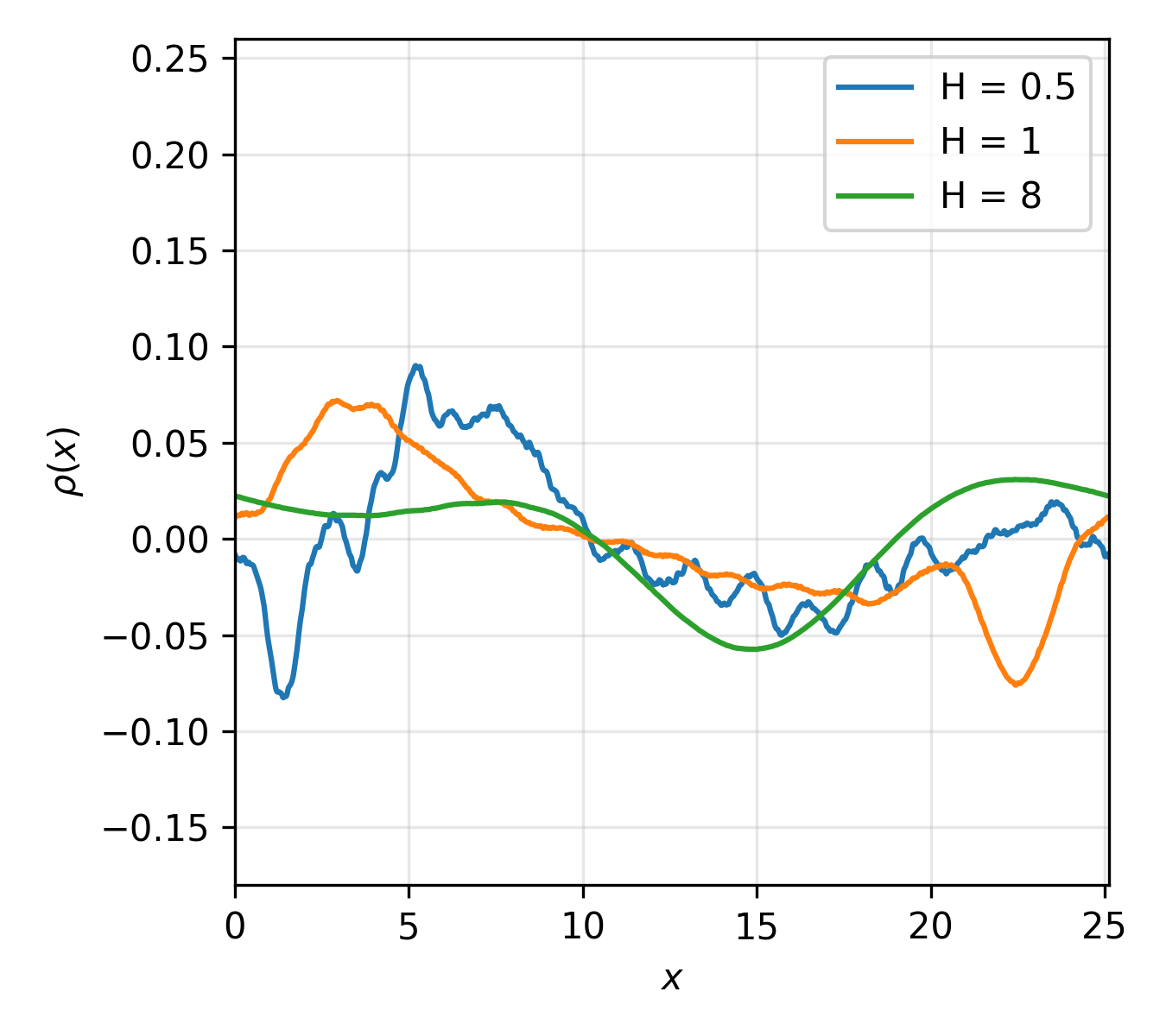}
        \caption{$t = 480$.}
        \label{fig:rho480}
    \end{subfigure} 
    \caption{Plasma densities at four different times for $H=0.5$, $H=1$, and $H=8$. we observe that the the smaller the $H$, the more complex structure it has. }
\end{figure*}

In HED plasmas, the plasma density is often one of the few quantities that can be measured as a function of space.  Here we look at the plasma density as a function of H (Figs.~\ref{fig:rho60}--\ref{fig:rho480}).  We observe that the plasma density could be studied as a possible finger print of how quantum the plasma is. Those with smaller H have less quantum diffraction and have a more complex structure.  Of course, to propose this as a good diagnostic, we would need to enhance the model to include collisional effects commensurate with these systems, as collisions can have a similar effect to larger H.  For now we note this as a plausible diagnostic to indicate how quantum the system is for laser driven plasmas.

\section{Conclusions}

We systematically explore how quantum diffraction and tunneling modify KEEN wave dynamics, and, by extension, the kinetic behavior of warm-dense plasmas.  By solving the one-dimensional Wigner–Poisson system, we tracked the evolution of a driven plasma over a wide span of the non-dimensional quantum parameter $H$.  The comparison with the classical limit reveals a coherent physical picture: as $H$ increases, diffraction adds an efficient phase-mixing mechanism that weakens particle trapping, damps the harmonic comb that characterizes classical KEEN waves, and drives the system toward stationarity on markedly shorter timescales.  At the modest value \(H\approx1\), the higher harmonics and phase-space vortices are visibly eroded; by \(H\simeq8\) the classical KEEN signature is reduced to a single-driver mode, with trapped structures washed out and density perturbations broadened.

As expected, increasing $H$ suppresses multiharmonic content and shortens relaxation times. Contrary to our initial expectation of a monotonic trend with H, we observe a small rise near $H \approx 1$ (consistent with selective harmonic locking). Because $H < 1$ requires a finer phase space resolution, the quantitative value has a larger uncertainty, so we report this feature as tentative. At $H=8$ we observe a pronounced early-time overshoot in electric energy is followed by a rapid relaxation to a nearly single-mode state, consistent with tunneling dominated phase mixing and a narrowed effective resonance.

These findings suggest that classical kinetic models may overestimate the longevity and energy content of subplasma frequency waves in warm-dense plasmas, which could in turn bias estimates of laser-plasma coupling and energy transport. Incorporating quantum-kinetic effects appears important for interpreting driven, nonequilibrium dynamics, particularly when diffraction is taken into consideration. Future extensions to include collisions, multi-species dynamics, and higher dimensionality should improve predictive modeling of nonlinear, driven kinetics in warm-dense plasmas.

\section*{Acknowledgments}

The authors acknowledge support from AFOSR grants FA9550-24-1-0254, DOE grant DE-SC0023164, ONR grant N00014-24-1-2242, and MSU Institute for Cyber-Enabled Research for their computer resources. Portions of this work were performed under the auspices of the U.S. Department of Energy by Lawrence Livermore National Laboratory under contract DE-AC52-07NA27344.

\end{document}